%% file: manuscript.tex
\documentclass[preprint,12pt]{elsarticle}

\usepackage{graphicx}
\usepackage{amsmath}
\usepackage{amssymb}
\usepackage{algorithm}
\usepackage{algpseudocode}

\usepackage{setspace}

\usepackage{lineno}

\makeatletter
\def\ps@pprintTitle{%
 \let\@oddhead\@empty
 \let\@evenhead\@empty
 \def\@oddfoot{}%
 \let\@evenfoot\@oddfoot}
\makeatother

 \biboptions{sort&compress}

\journal{}

\begin{document}

\begin{frontmatter}

\title{Spatial Variation of the Correlated Color Temperature of Lightning Channel}

\author{Nobuaki Shimoji\corref{cor1}}
\ead{nshimoji@tec.u-ryukyu.ac.jp}
\cortext[cor1]{Corresponding author}
\author{Ryoma Aoyama}

\address{Department of Electrical and Electronics Engineering, University of the Ryukyus, 1 Senbaru, Nishihara, Okinawa, 903-0213, Japan}

\begin{abstract}
  In present work, we propose the analysis method of lightning based on the color analysis.
  We analyzed the digital still images in which the cloud-to-ground (CG) and intracloud (IC) lightning flashes are shown.
  Applying some digital image processing techniques, we extracted lightning channels.
  Then, the correlated color temperature (CCT) of the extracted lightning channels was obtained by mapping digital pixels of the extracted lightning channels to CIE 1931 $xy$-chromaticity diagram.
  Our results indicate that the CCT of lightning channels changes spatially.
  Furthermore, it suggests that the energy of lightning channels changes spatially.

\end{abstract}


\end{frontmatter}

\input{sec_01}

\input{sec_02}

\input{sec_03}

\input{sec_04}

\input{sec_05}

\input{sec_06}

\input{sec_07}

\bibliographystyle{elsarticle-num}
\bibliography{bibtex_cct_lightning}

\end{document}

%% file: sec_01.tex
\section{Introduction}
\label{introduction}


  Lightning is very high-speed and very large natural phenomenon.
  In general, we cannot predicate when and where it occurs.
  Also, it is difficult that we directly observe properties of lightning.
  From these reason, lightning is indirectly observed by using a electric field sensor, a optical sensor, or digital still/video camera etc.
  It can be considered that from now on, performance of these instruments recording lightning will be higher.
  In recent years, the studies of lightning using the high-performance and high speed camera have increased~\cite{Tom_A_Warner,Tom_A_Warner_2,J_Montanya}.
  These results provide new insights to lightning researchers.
  In addition, it is also considered that developing the analytical methods is necessary.
  In this paper we propose the new analytical method of lightning based on the color analysis technique.

  We analyzed the lightning channel based on the color analysis.
  First of all, we extracted lightning channels in digital still images and then performed the color analysis of the extracted lightning channels, where we used OpenCV library~\cite{OpenCV}.
  Our results show that the CCT of lightning channels changes spatially.
  It is considered that if the CCT of a lightning channel is higher than other channels, the energy of the lightning channel also higher than the other.
  Thus, since the CCT of lightning channels changes spatially, the energy of lightning also changes spatially.

%% file: sec_02.tex
\section{Lightning images and extraction of lightning channel}
\label{sec:lightning_images_and_extraction_of_lightning_channel}

\subsection{Lightning images}
\label{subsec:Lightning images}

We used two lightning flash images shown in Fig.~\ref{fig:lightning images}~\cite{YutakaAoki}.
The images were captured in Chikusei City, Ibaraki Prefecture, Japan, on October 27th, 2008.
The images were saved as uncompressed files and optical filters, e.g. ND filters, did not be used.
For this reason, it is considered that the information deteriotation on the digital images is minimized.
Therefore the lightning flash images used in this work can be used to analyze colors of the lightning channels.

\subsection{Extraction of lightning channels}
\label{subsec:Extraction of lightning channels}

\begin{figure}
  \begin{center}
    \includegraphics[width=40.0mm]{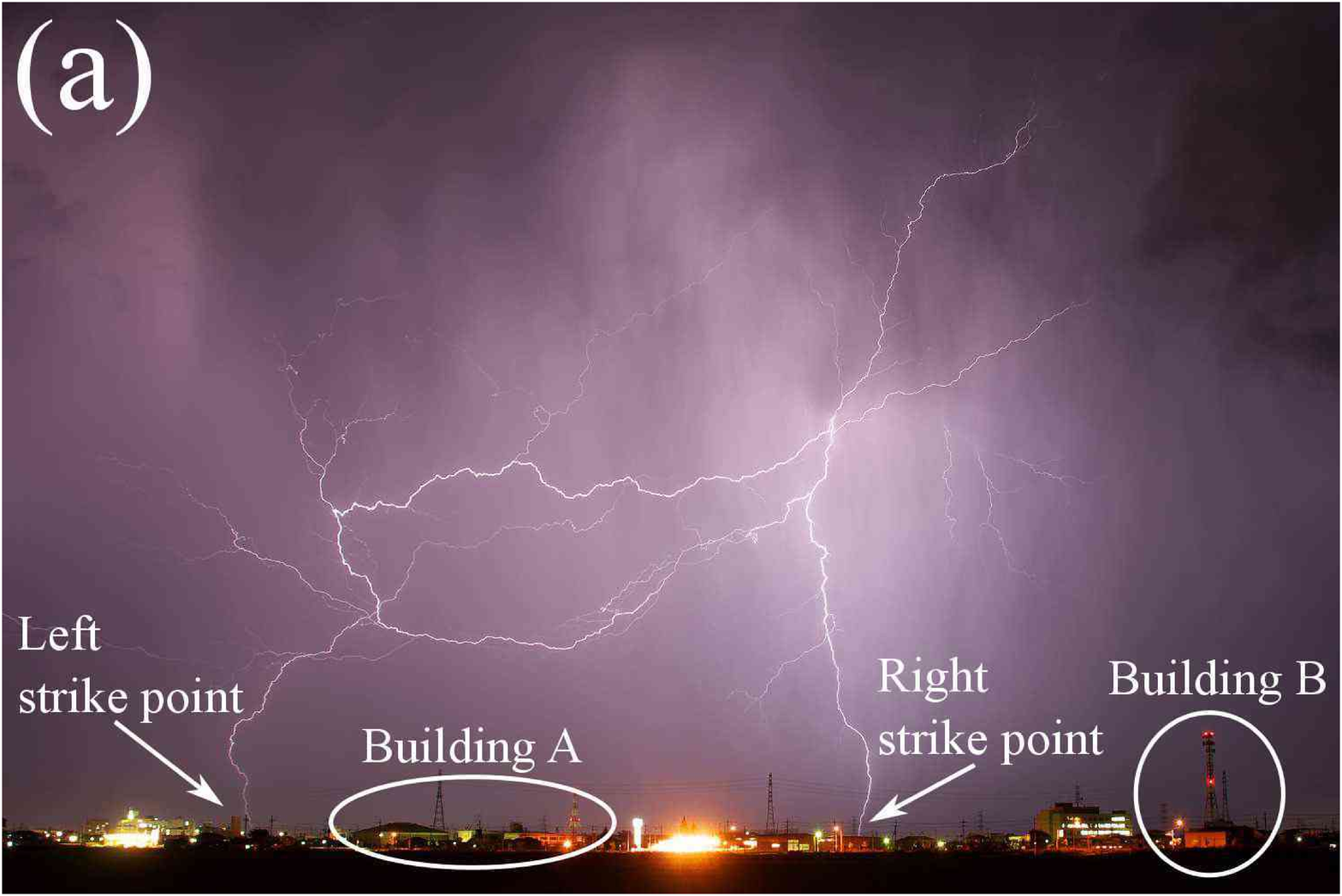}
    \includegraphics[width=40.0mm]{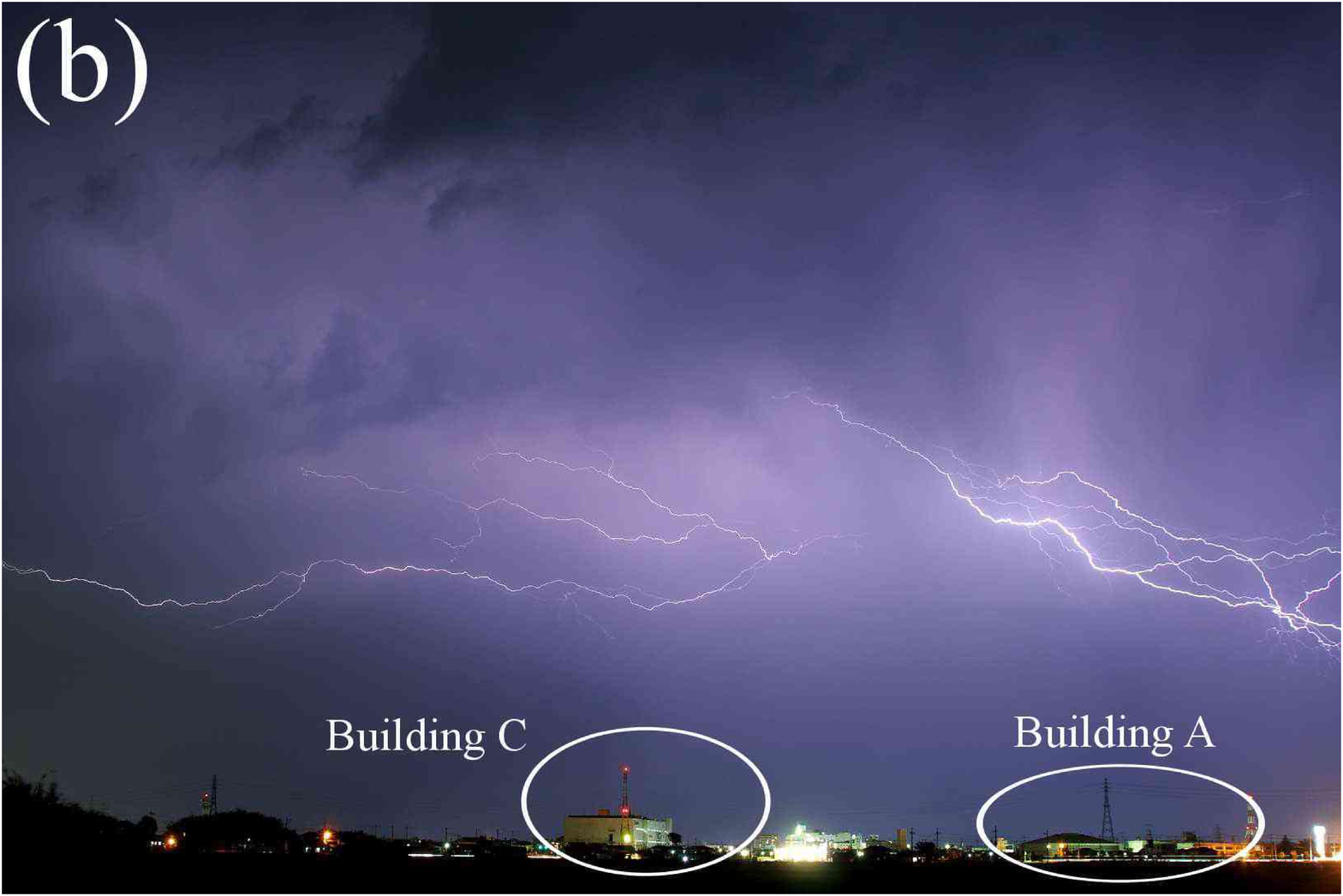}
    \caption{(Left) CG and (Right) IC lightning images provided by storm chaser Yutaka Aoki. There are the landmarks, building A (Shimodate-Minami Junior High School), building B (Shimodate River Office), and building C (Nippon Meat Packers, Inc.) in the images. These landmarks were used to define the directions of strike points in the left figure and the fields of view of both the images.}
    \label{fig:lightning images}
  \end{center}
\end{figure}

\begin{figure*}[hbtp]
  \begin{center}
    \includegraphics[width=40.0mm]{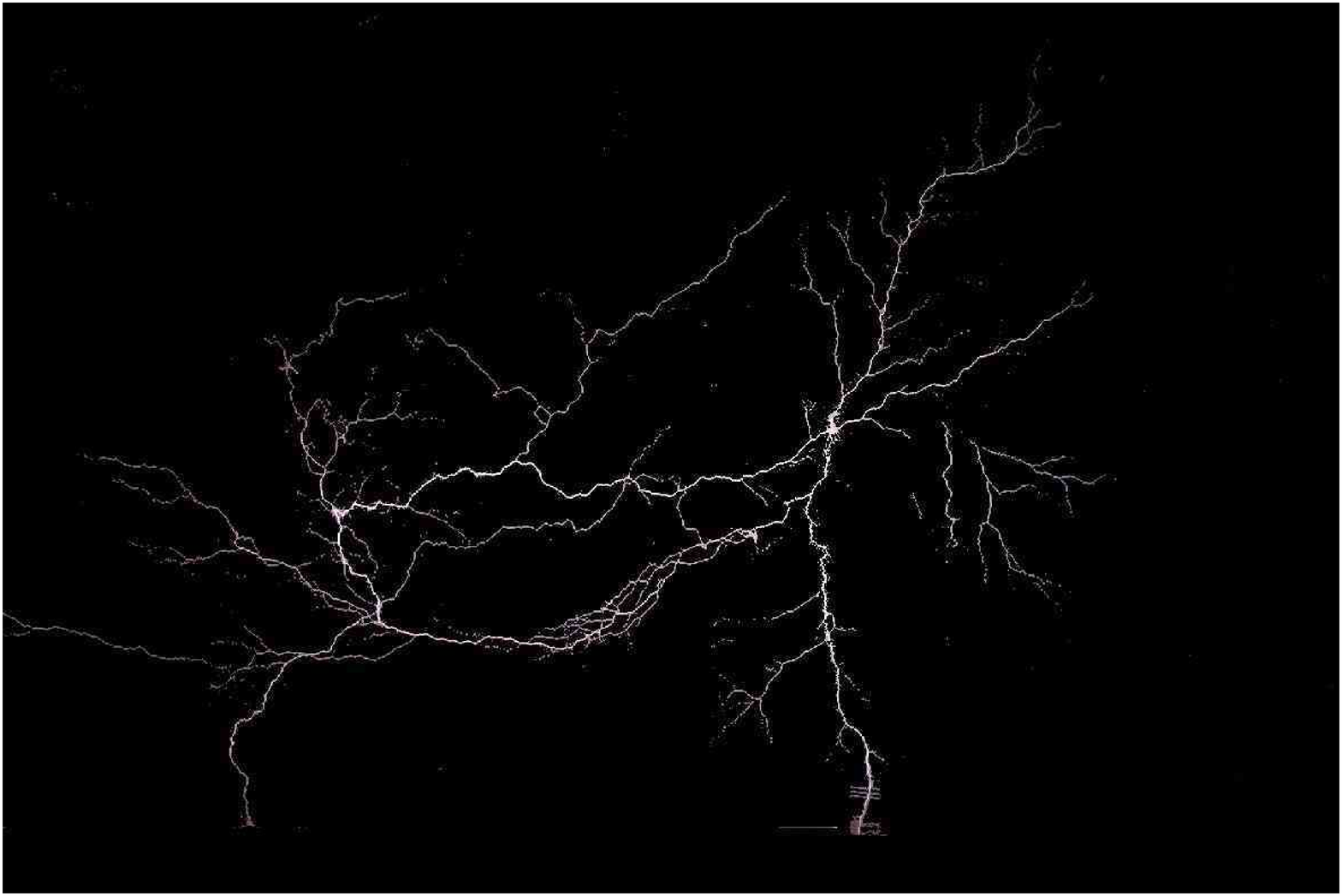}
    \includegraphics[width=40.0mm]{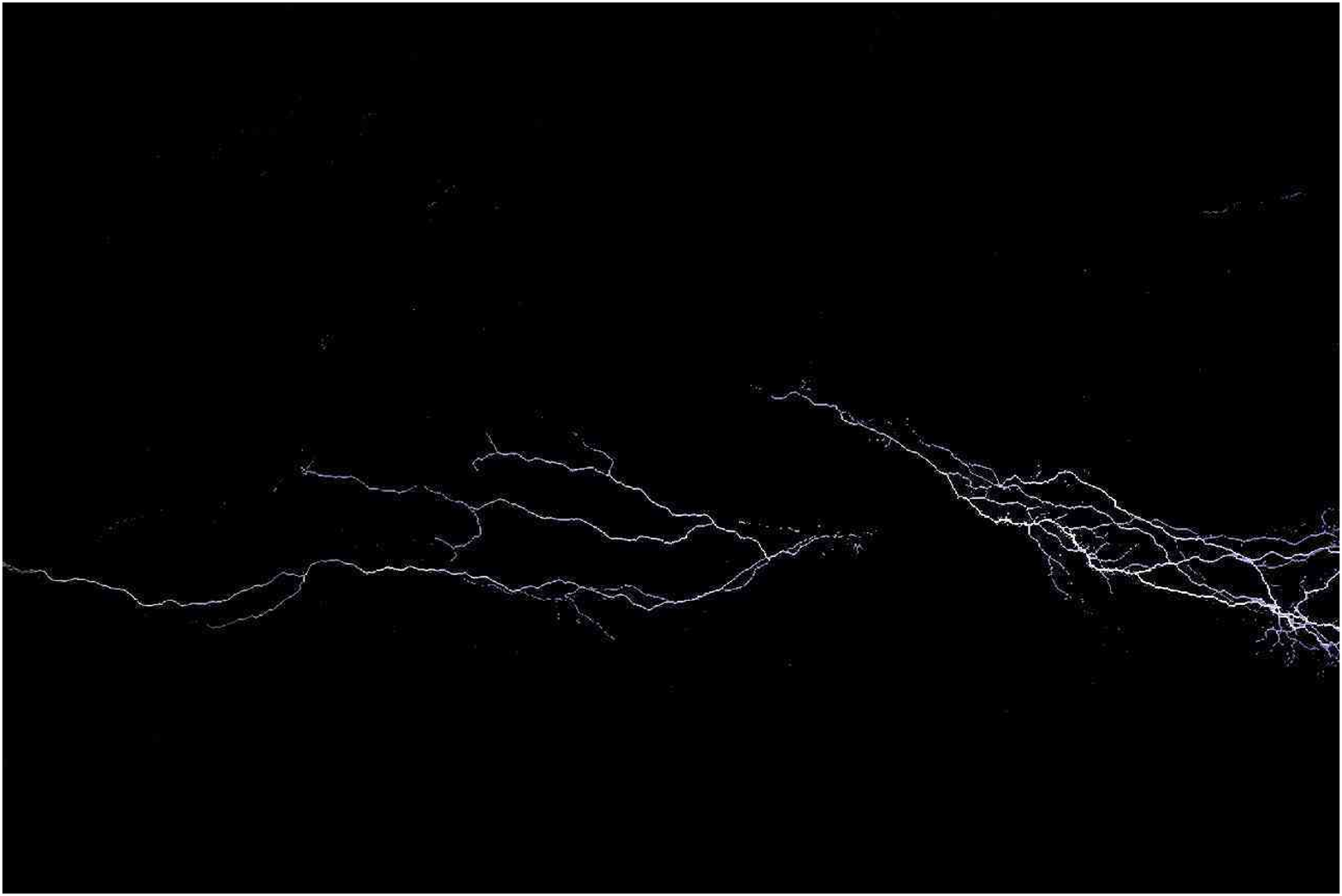}
    \caption{Extracted lightning channels for the (Left) CG and (Right) IC flash images which were obtained applying digital image processing tequniques to Fig.~\ref{fig:lightning images} (Left) and (Right), respectively.}
    \label{fig:masking images}
  \end{center}
\end{figure*}

We have to extract lighting channels from the lightning flash images (Fig.~\ref{fig:lightning images}).
We can see many unwanted parts for the color analysis (e.g. buildings, artificial light sources, and cloud luminescence accompany with a lightning flash in the images.
These unwanted parts are unnecessary to analyze colors of the lightning channels.
Thus we extracted the lightning channels in the images by applying digital image processing techniques.
The extracted lightning channels are shown in Fig.~\ref{fig:masking images}.

The extraction method is based on masking of digital image processing.
The extraction procedure is as follows: (1)Inpainting, (2)Smoothing (blur), (3)Sobel edge detection, (4)Thresholding, (5)Smoothing (median filter), (6)Filling process, (7)Thinning process, and (8)Masking.
We run in turn the image processing above using OpenCV library~\cite{OpenCV}.
Steps (6) and (7) in the extraction procedure are created by our work.

%% file: sec_03.tex
\section{Reduction technique of chromatic aberration}
\label{sec:reduction technique of chromatic aberration}

\begin{figure}[hbtp]
  \begin{center}
    \includegraphics[height=50.0mm]{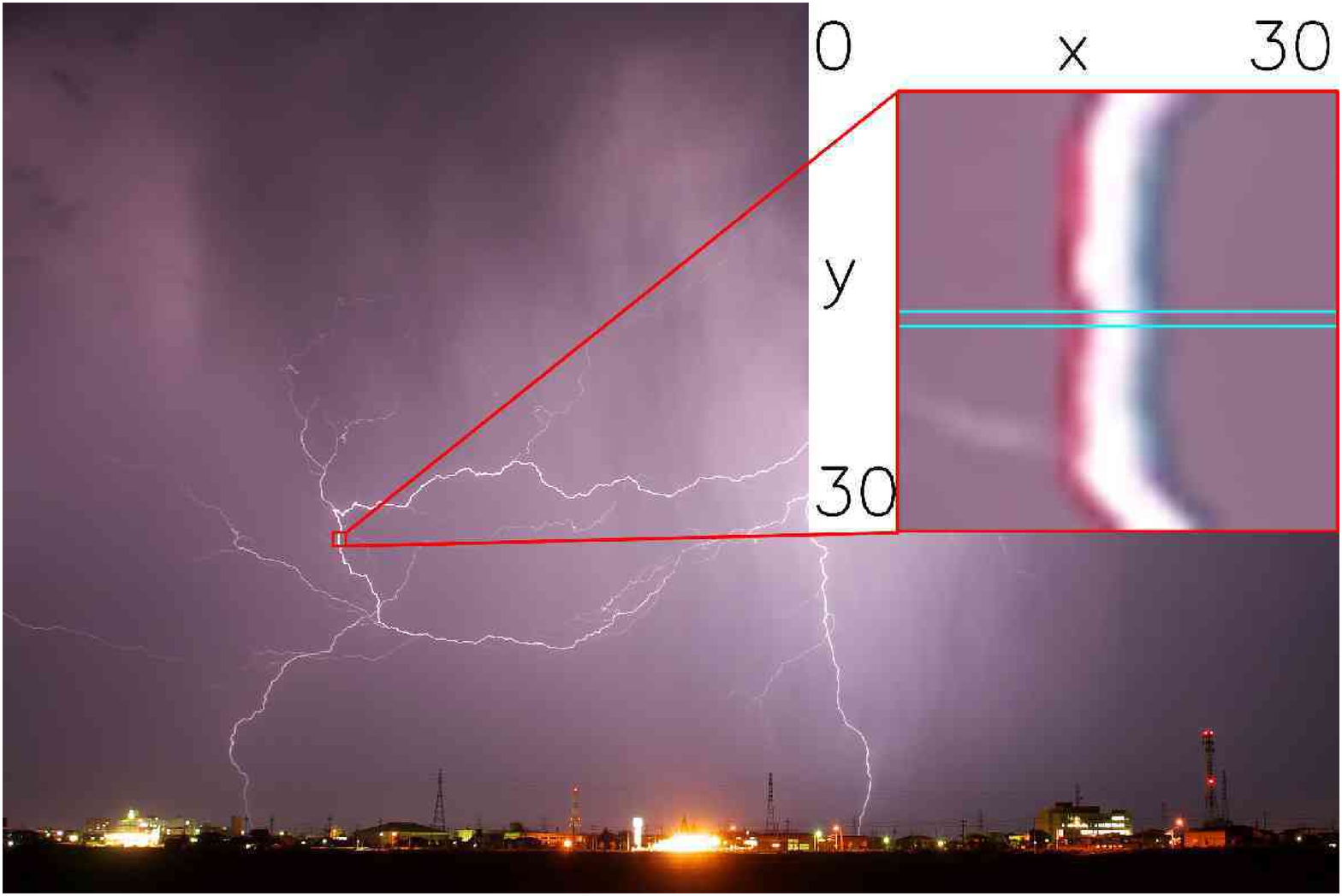}
    \includegraphics[width=60.0mm]{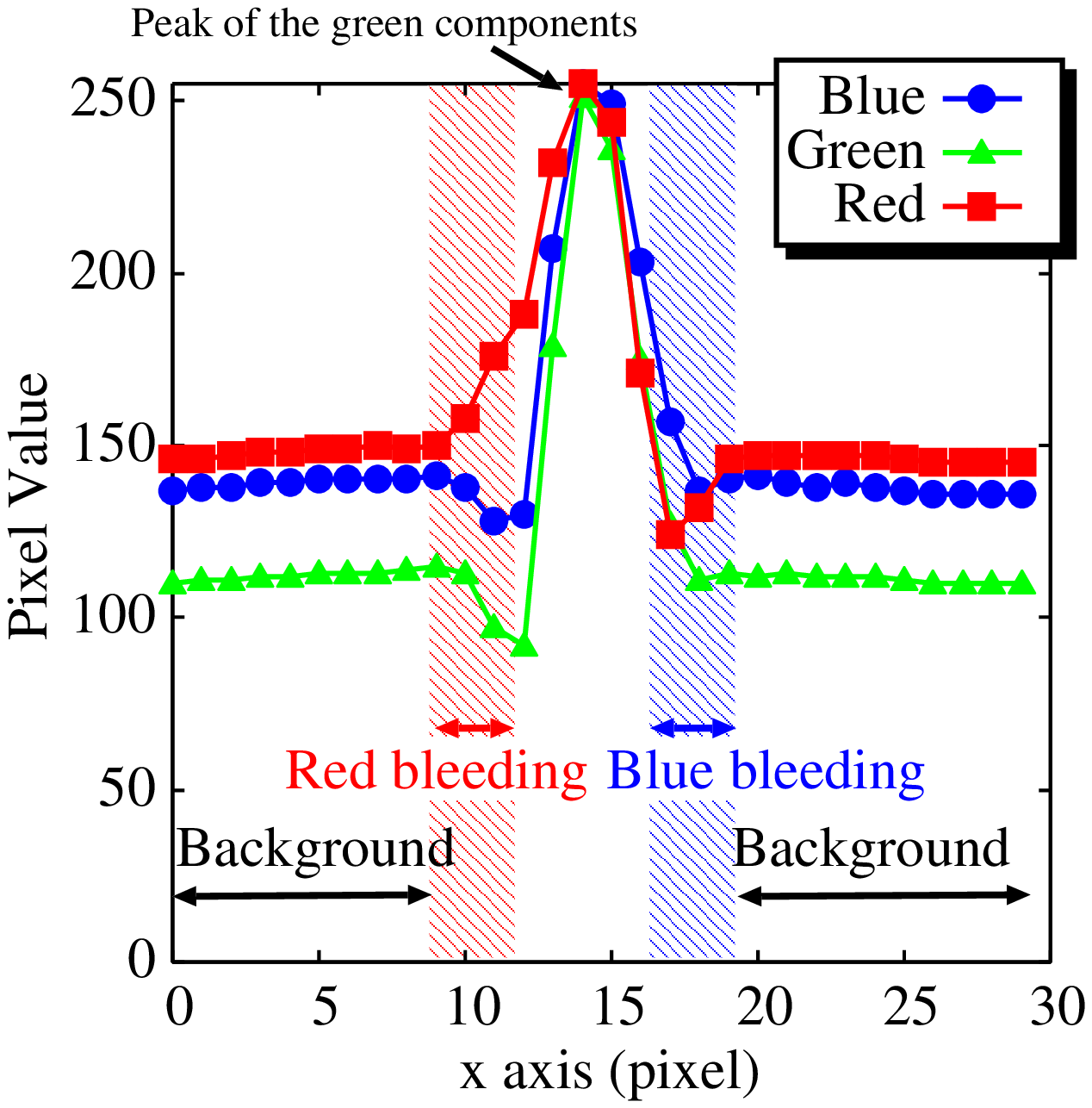}
    \caption{(Left) Original CG flash image and the insets which is the enlarged figure of the lightning channel. The size of the enlarged figure is $30 \times 30$ pixels. (Right) Pixel values for the RGB components ($R$, $G$, $B$) along with the line between two pale blue lines in the inset. The red solid square, green solid triangle, and blue solid circle denote the pixel value of R-, G-, and B-components, respectively.}
    \label{fig:chromatic aberration CG}
  \end{center}
\end{figure}

\begin{figure}[hbtp]
  \begin{center}
    \includegraphics[height=50.0mm]{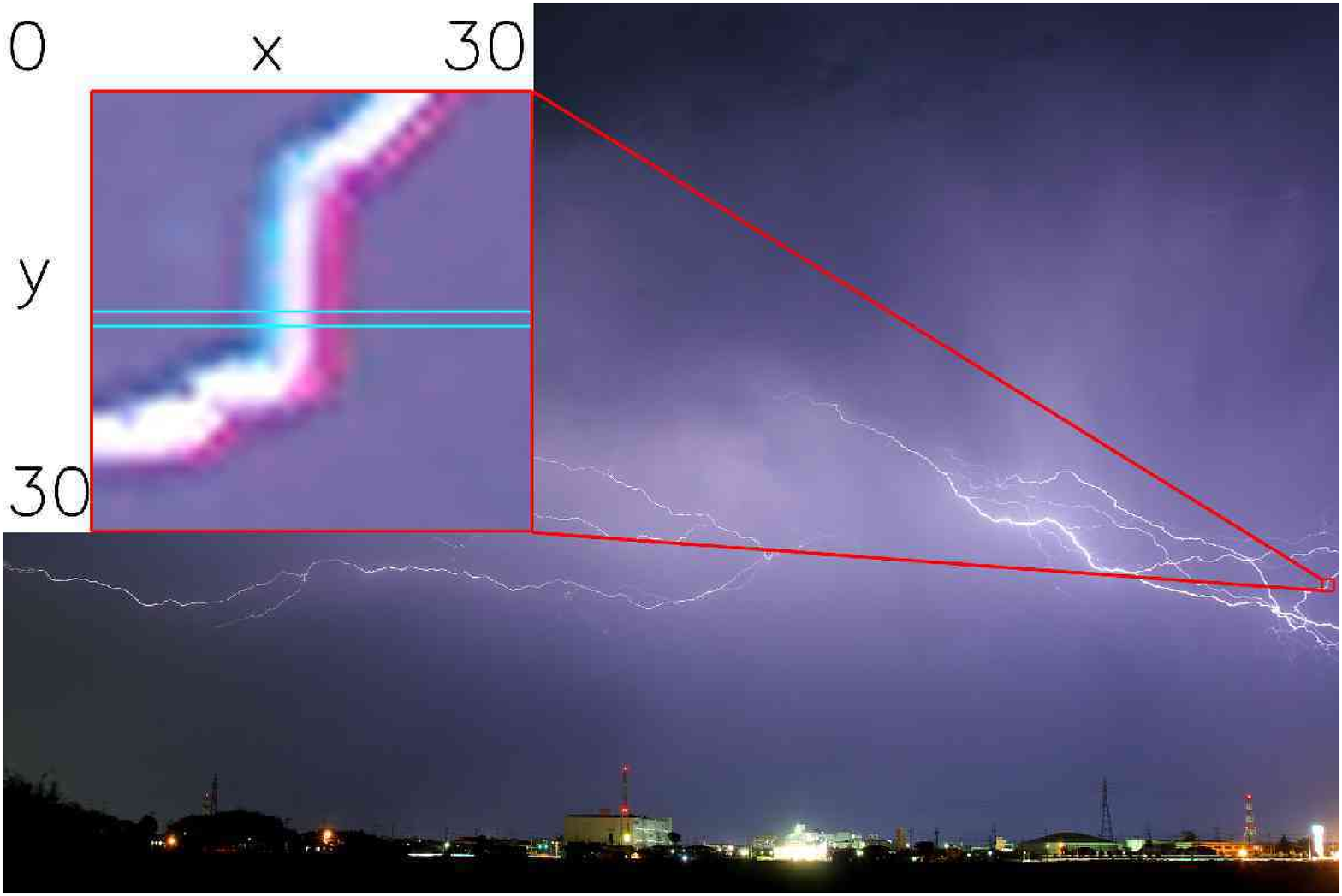}
    \includegraphics[width=60.0mm]{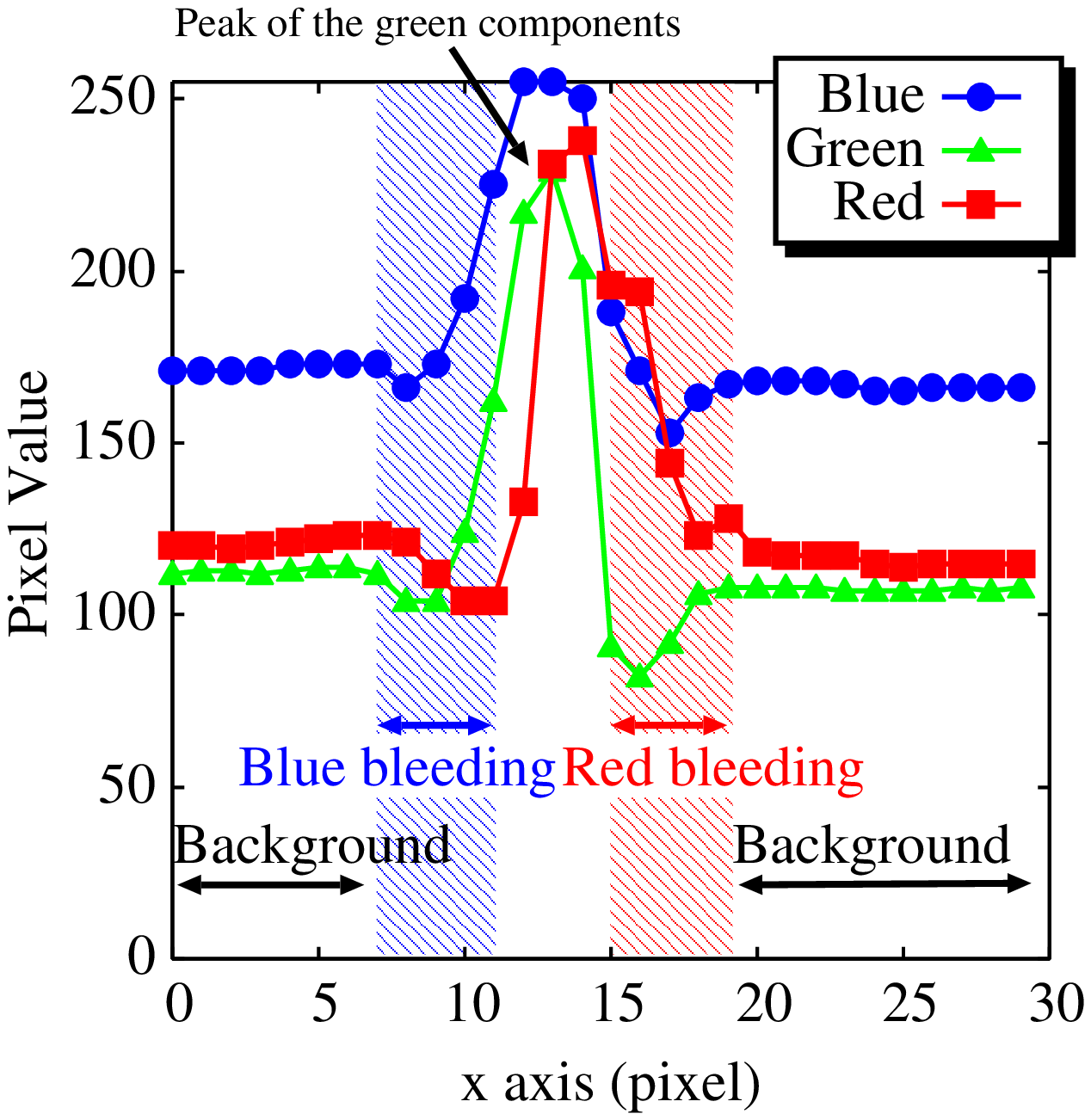}
    \caption{(Left) Original IC flash image and the insets which is the enlarged figure of the lightning channel. The size of the enlarged figure is $30 \times 30$ pixels. (Right) Pixel values for RGB components ($R$, $G$, $B$) along with the line between two pale blue lines in the inset. The red solid square, green solid triangle, and blue solid circle denote the pixel value of R-, G-, and B-components, respectively.}
    \label{fig:chromatic aberration IC}
  \end{center}
\end{figure}

\begin{figure}[hbtp]
  \begin{center}
    \includegraphics[width=60.0mm]{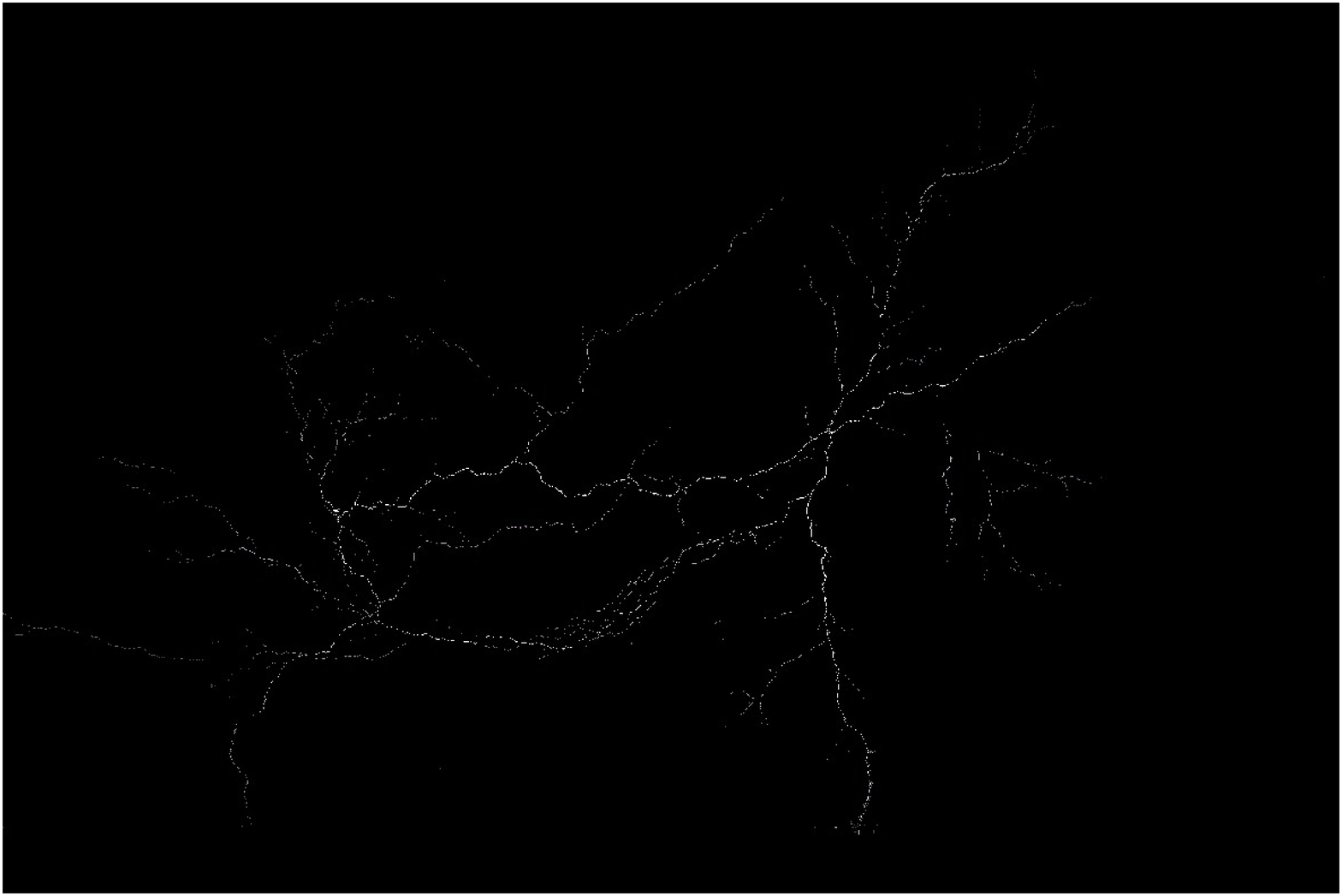}
    \includegraphics[width=60.0mm]{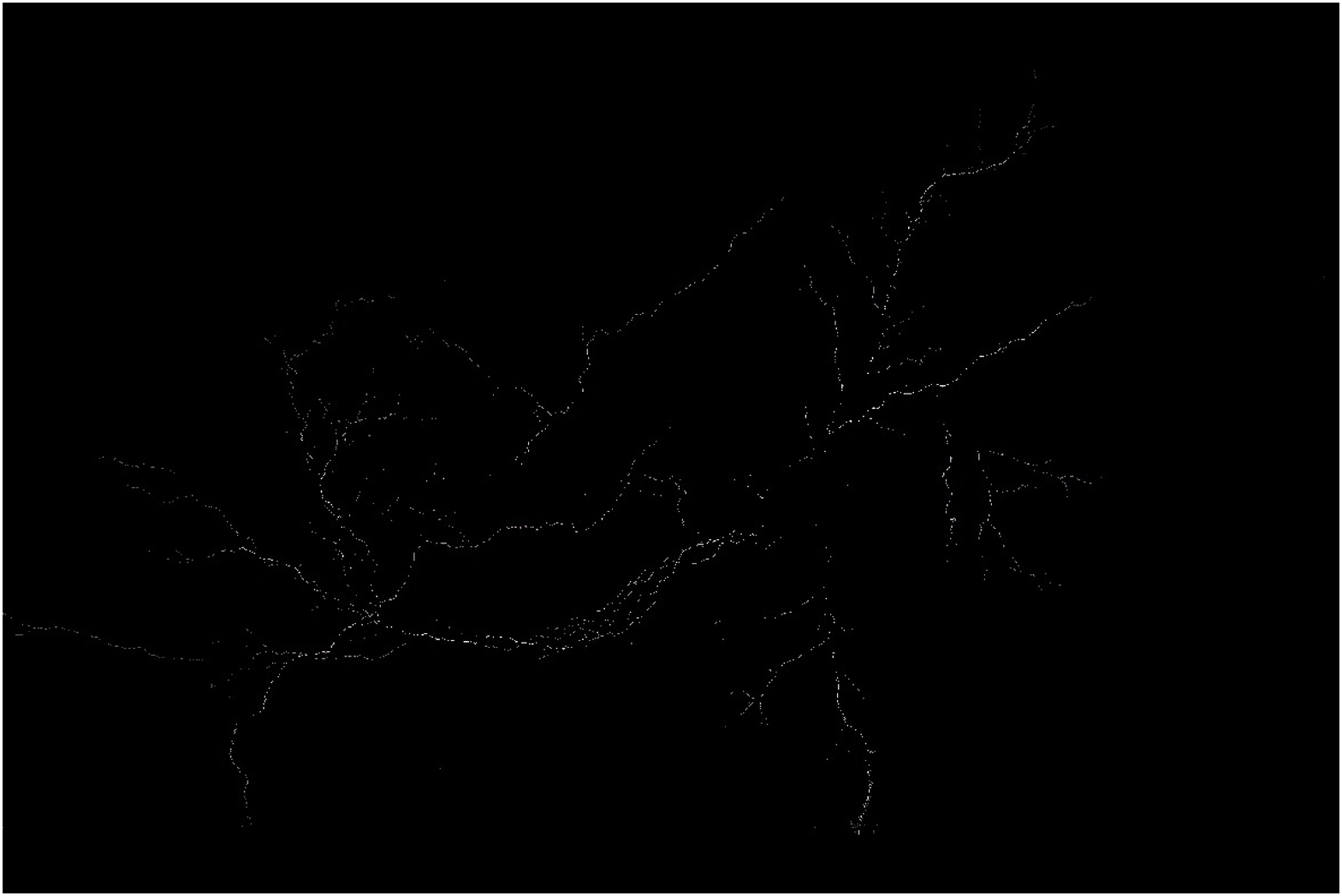}
    \caption{CG lightning channels applied Algorithm~\ref{algorithm:reduction algorithm} to the images Fig.~\ref{fig:masking images} (a). The left side figure containes the saturated pixels reaching $255$, but the right side does not contain it.}
    \label{fig:CG extraction}
  \end{center}
\end{figure}

\begin{figure}[hbtp]
  \begin{center}
    \includegraphics[width=60.0mm]{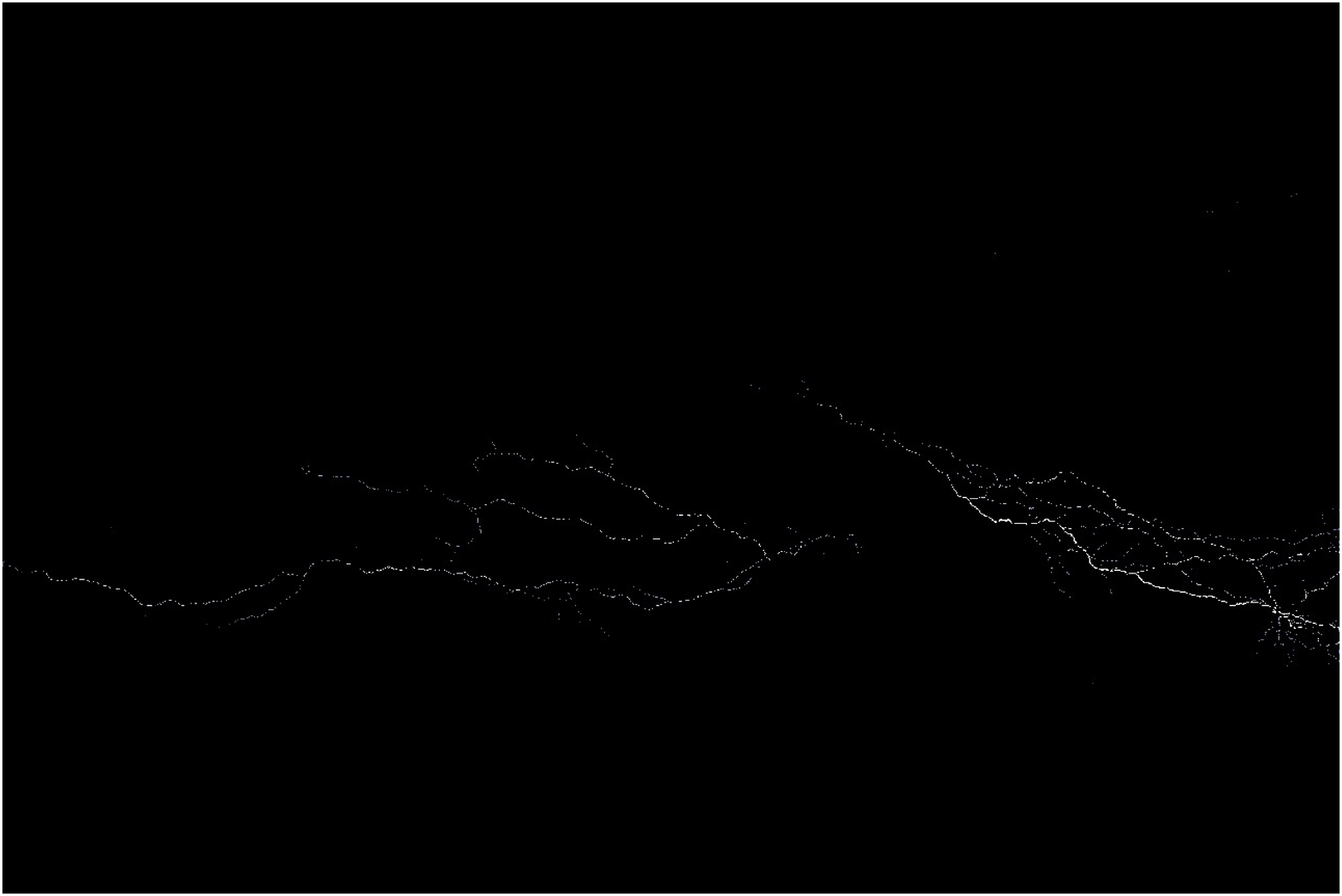}
    \includegraphics[width=60.0mm]{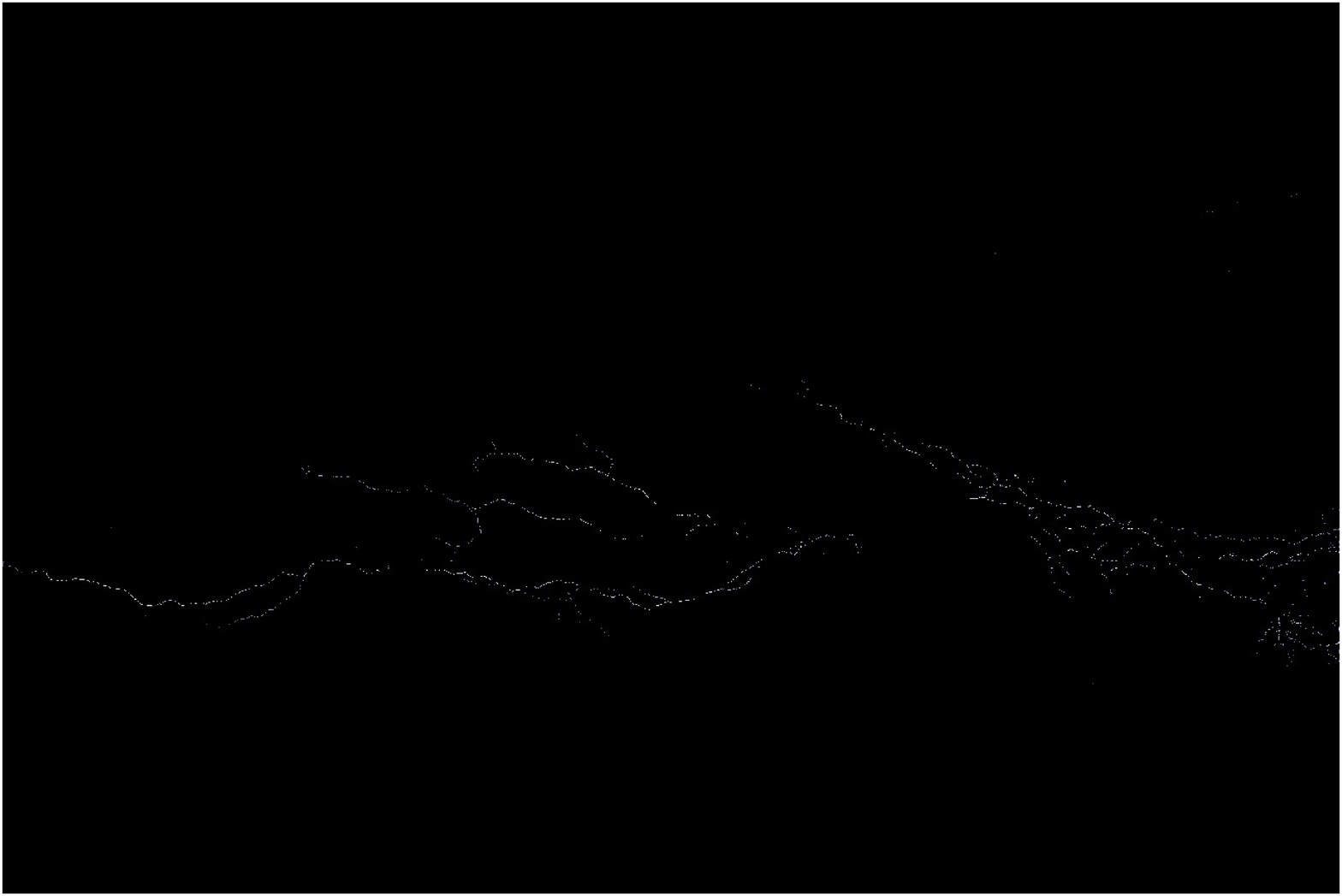}
    \caption{IC lightning channels applied Algorithm~\ref{algorithm:reduction algorithm} to the images Fig.~\ref{fig:masking images} (b). The left side figure containes the saturated pixels reaching $255$, but the right side does not contain it.}
    \label{fig:IC extraction}
  \end{center}
\end{figure}

In this section we explain a reduction tequnique of chromatic aberration in the lightning flash images.
Usually, in a image created by the optical system, chromatic aberration is unavoidable.
When we analyze lightning channels on digital images, the chromatic aberration should be removed from the images.
Figs.~\ref{fig:chromatic aberration CG} and \ref{fig:chromatic aberration IC} show the example of the chromatic aberration on the lightning channel in the CG and IC flash images.
The insets on Figs.~\ref{fig:chromatic aberration CG} and \ref{fig:chromatic aberration IC} show the enlarged lightning channel, which have the size $30 \times 30$ pixels.
Along the lightning channels in the insets (Figs.~\ref{fig:chromatic aberration CG} and \ref{fig:chromatic aberration IC}), the red and blue color is running.
This spread is the chromatic aberration.
In order to see the chromatic aberration in detail, as an example we plot the pixel value on the line between two pale blue lines in the insets (see Figs.~\ref{fig:chromatic aberration CG} (Right) and \ref{fig:chromatic aberration IC} (Right)).
In Figs.~\ref{fig:chromatic aberration CG} (Right) and \ref{fig:chromatic aberration IC} (Right), the vertical axis denotes the pixel value which have the range $0$ -- $255$ ($256$ levels), and the $x$-axis denotes the line between the two pale blue lines.
From the variation of the pixel values, we can see that there is the difference of the spread of the pixel value for the R-, G-, and B-components.
This spreads attribut to the chromatic aberration.
The chromatic aberration is unnecessary to analyze the color of the lightning channel.
Usually, in optics, it is well known that the green light has a smaller dispersion of the chromatic aberration than the red and blue light.
This fact indicates that, in Figs.~\ref{fig:chromatic aberration CG} (Right) and \ref{fig:chromatic aberration IC} (Right), the effect of the chromatic aberration near the peak of the green is smaller.
Hence, in order to eliminate the chromatic aberration, we created Algorithm~\ref{algorithm:reduction algorithm}.
\begin{algorithm}
\caption{Reduction algorithm on the chromatic aberration.}
\label{algorithm:reduction algorithm}
\begin{algorithmic}
\State // $(x, y)$: coordinate in the image.
\State // $Red(x, y)$: Pixel value of the red component at the $(x, y)$ coordinate.
\State // $Green(x, y)$: Pixel value of the green component at the $(x, y)$ coordinate.
\State // $Blue(x, y)$: Pixel value of the blue component at the $(x, y)$ coordinate.
\State // $RGB(x, y)$: Pixel value consisting of $Red$, $Green$, $Blue$ at the $(x, y)$ coordinate.
\For{$y = 2$ to $\text{image height} - 2$}
\For{$x = 2$ to $\text{image width} - 2$}
\If{$Green(x - 2, y) < Green(x - 1, y) < Green(x, y) > Green(x + 1, y) > Green(x + 2, y)$}
\State $RGB(x, y) \leftarrow ( Red(x, y), Green(x, y), Blue(x, y) )$
\Else
\State $RGB(x, y) \leftarrow ( 0, 0, 0 )$
\EndIf
\EndFor
\EndFor
\end{algorithmic}
\end{algorithm}
This algorithm extract the RGB components ($R$, $G$, $B$) at the peak of the G-component of lightning channel, namely, the RGB conponents at the part of the smaller effect of the chromatic aberration.
Figs.~\ref{fig:CG extraction} and \ref{fig:IC extraction} show images applied the Algorithm~\ref{algorithm:reduction algorithm} to the extracted lightning channels shown in Fig.~\ref{fig:masking images} (a) and (b).
The left side of Figs.~\ref{fig:CG extraction} and \ref{fig:IC extraction} containes the saturated pixels reaching $255$, but the right side of Figs.~\ref{fig:CG extraction} and \ref{fig:IC extraction} does not contain it.

%% file: sec_04.tex
\section{Variation of the correlated color temperature of the lightning channel}
\label{sec:variation of the correlated color temperature of the lightning channel}

We have studied the variation of the CCT of the lightning channels shown in Figs.~\ref{fig:CG extraction} and \ref{fig:IC extraction}.
As indicated by Uman and Orville ~\cite{Uman_Orville}, lightning opacity is thin.
That is, lightning is not the Planckian radiator.
Therefore, we cannot identify the temperature of the lightning channel by considering the color temperature.
However the color of the light emitted from the lightning is meaningful, since the color contains the information in the interior of lightning channels.
Thus we draw attention to the CCT of the lightning channels.

\subsection{Correlated color temperature}
\label{ssec:correlated color temperature}

\begin{figure}[hbtp]
  \begin{center}
    \includegraphics[width=40.0mm]{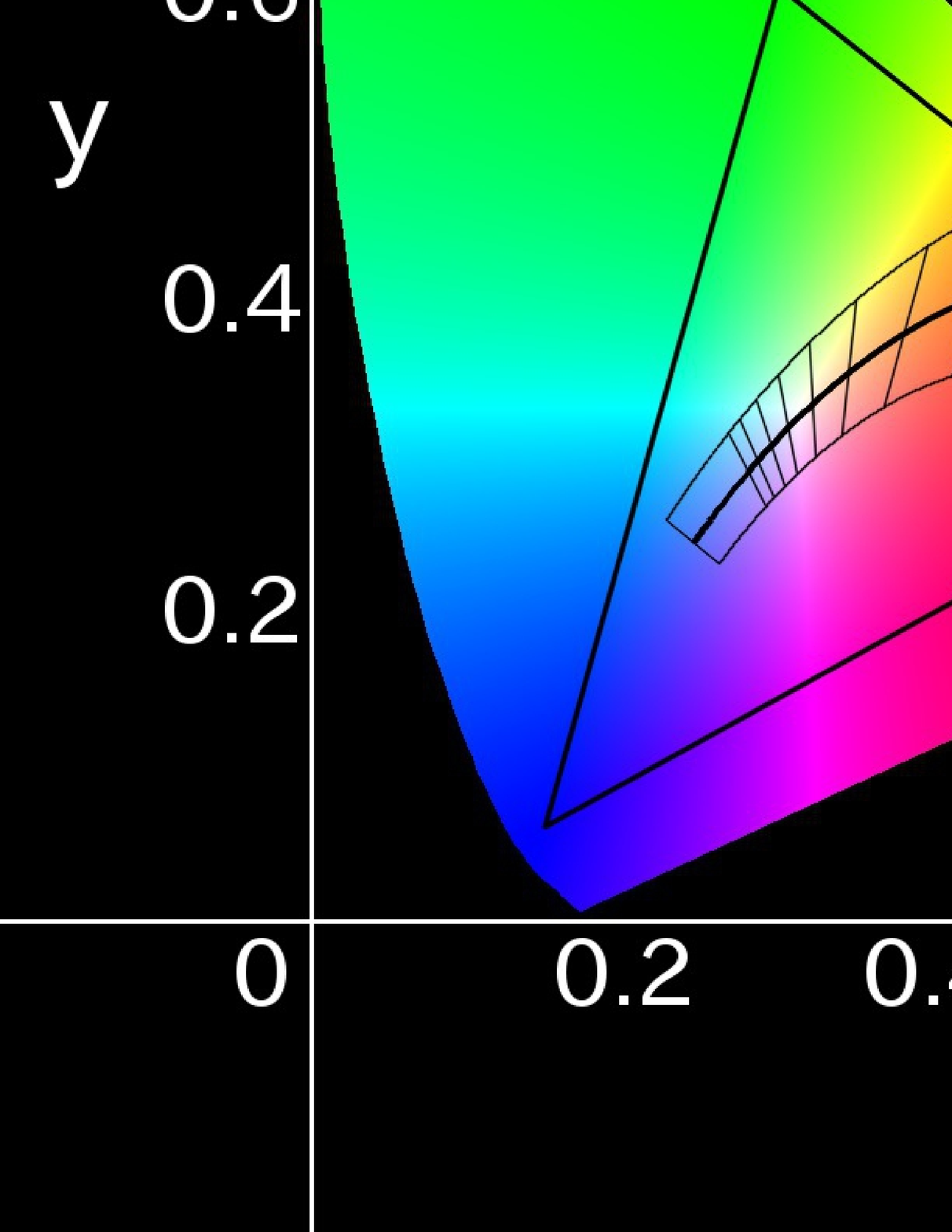}
    \includegraphics[width=40.0mm]{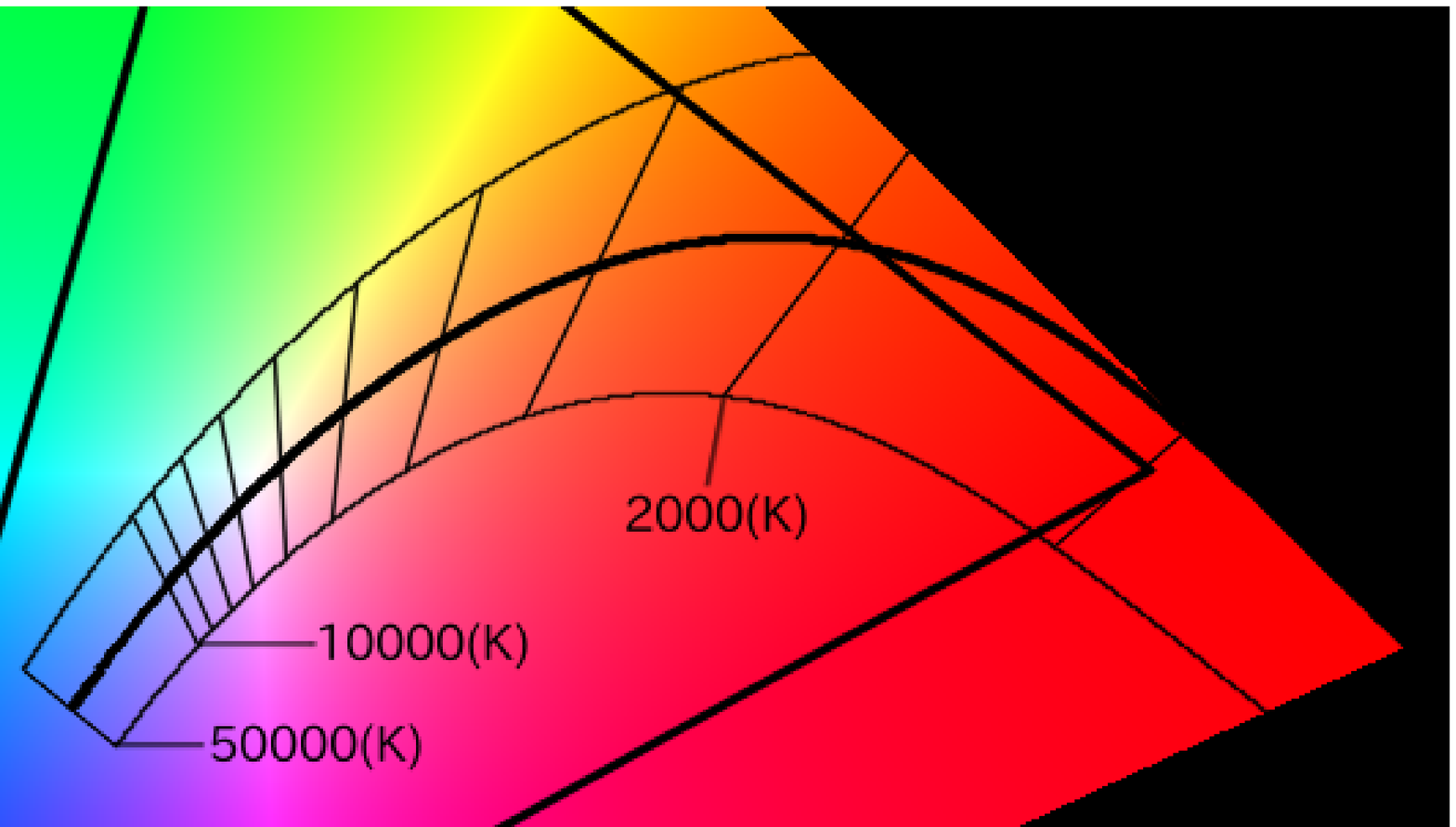}
    \caption{(Left) CIE 1931 $xy$-chromaticity diagram in which the sRGB color triangle and the Planckian locus are depicted. (Right) enlarged figure near the Planckian locus where the thick line denotes the Planckian locus and the thin lines denote the isotemperature lines. The length of the isotemperature lines is $\pm 0.02\Delta$uv.}
    \label{fig:cie 1931 xy chromaticity diagram}
  \end{center}
\end{figure}

We describe conversion methods from sRGB color space to the CIE XYZ color space.
The color space of the lightning flash images (Fig.~\ref{fig:lightning images}) is sRGB\cite{IEC}.
The CCT is discussed on the $xy$-chromaticity diagram.
Fig.~\ref{fig:cie 1931 xy chromaticity diagram} (left) shows the CIE 1931 $xy$-chromaticity diagram in which the sRGB color triangle and the Planckian locus are depicted.
Fig.~\ref{fig:cie 1931 xy chromaticity diagram} (right) shows a enlarged figure near the Planckian locus in which the Planckian locus (thick line), the equal $\Delta$uv lines of $\pm 0.02\Delta$uv along with the Planckian locus, and the isotemperature lines (thin lines) acrossing with the Planckian locus are depicted.
The length of the isotemperature lines is $\pm 0.02\Delta$uv.
The CCT is generally discussed in the range $\pm 0.02\Delta$uv.

To analyze the lightning channels (Figs.~\ref{fig:CG extraction} and \ref{fig:IC extraction}) based on the CCT, we have converted the sRGB color space to the CIE XYZ color space.
Here are the conversion method from the sRGB color space to the CIE XYZ color space~\cite{IEC}.

We first divid the nonlinear $sRGB$ components, $R_{sRGB}$, $G_{sRGB}$ and $B_{sRGB}$, by $255$ and then we obtain the normalized values,
\begin{eqnarray*}
  R_{sRGB}^{\prime} &=& R_{sRGB} / 255,\\
  G_{sRGB}^{\prime} &=& G_{sRGB} / 255,\\
  B_{sRGB}^{\prime} &=& B_{sRGB} / 255.
\end{eqnarray*}
Then the normalized nonlinear $sRGB$ components, $R_{sRGB}^{\prime}$, $G_{sRGB}^{\prime}$ and $B_{sRGB}^{\prime}$, are transformed to linear $sRGB$ components, $R_{linear}$, $G_{linear}$ and $B_{linear}$, as follows :

If $R_{sRGB}^{\prime}$, $G_{sRGB}^{\prime}$, $B_{sRGB}^{\prime} \le 0.04045$, then
\begin{eqnarray*}
  R_{linear} &=& R_{sRGB}^{\prime} / 12.92 , \\
  G_{linear} &=& G_{sRGB}^{\prime} / 12.92 , \\
  B_{linear} &=& B_{sRGB}^{\prime} / 12.92 , \\
\end{eqnarray*}
else if $R_{sRGB}^{\prime}$, $G_{sRGB}^{\prime}$, $B_{sRGB}^{\prime} > 0.04045$,
\begin{eqnarray*}
  R_{linear} &=& \left( \left( R_{sRGB}^{\prime} + 0.055 \right) / 1.055 \right)^{2.4}, \\
  G_{linear} &=& \left( \left( G_{sRGB}^{\prime} + 0.055 \right) / 1.055 \right)^{2.4}, \\
  B_{linear} &=& \left( \left( B_{sRGB}^{\prime} + 0.055 \right) / 1.055 \right)^{2.4}. \\
\end{eqnarray*}
The linear $sRGB$ components $R_{linear}$, $G_{linear}$ and $B_{linear}$ are converted to the CIE XYZ system by the equation~\cite{ITU}:
\begin{eqnarray}
  \begin{bmatrix}
    X \\
    Y \\
    Z
  \end{bmatrix}
  =
  \begin{bmatrix}
    0.4124 & 0.3576 & 0.1805 \\
    0.2126 & 0.7152 & 0.0722 \\
    0.0193 & 0.1192 & 0.9505
  \end{bmatrix}
  \begin{bmatrix}
    R_{linear} \\
    G_{linear} \\
    B_{linear}
  \end{bmatrix}
  .
  \label{eqn:RGB-XYZ}
\end{eqnarray}
Using the tristimulus value $X, Y,$ and $Z$ the chromaticity coordinates $x, y$ on the $xy$-chromaticity diagram are expressed by
\begin{eqnarray}
  x = \frac{X}{X + Y + Z} ,
  y = \frac{Y}{X + Y + Z} .
\end{eqnarray}

\subsection{Correlated color temperature of the lightning channels}
\label{ssec:correlated color temperature of lightning channels}

\begin{figure}[hbtp]
  \begin{center}
    \includegraphics[width=60.0mm]{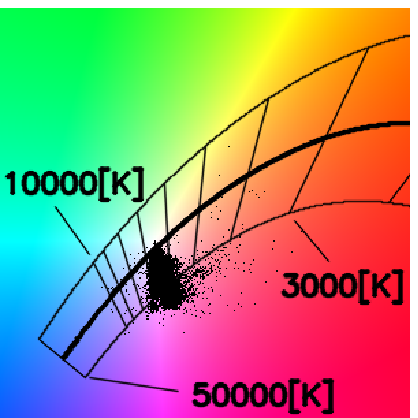}
    \includegraphics[width=60.0mm]{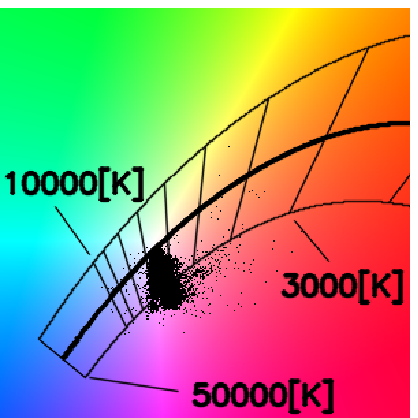}
    \caption{All projected points for the CG lightning channel shown in Fig.~\ref{fig:CG extraction}. The thick line denotes the Planckian locus. The thin lines crossing the Planckian locus are the isotemperature lines. The thin lines along the Planckian locus are the equal $\Delta$uv lines of $\pm 0.02\Delta$uv. The left side containes the saturated pixels reaching $255$, but the right side does not contain.}
    \label{fig:CG xy chromaticity diagram}
  \end{center}
\end{figure}

\begin{figure}[hbtp]
  \begin{center}
    \includegraphics[width=60.0mm]{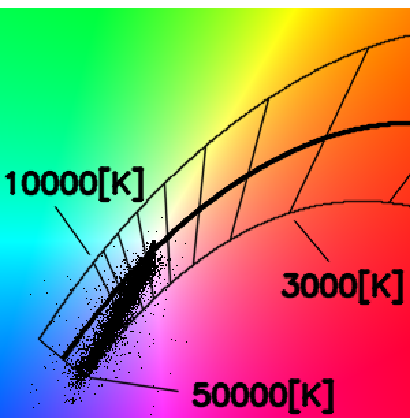}
    \includegraphics[width=60.0mm]{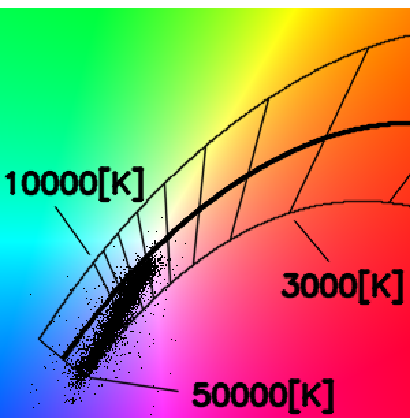}
    \caption{All projected points for the IC lightning channel shown in Fig.~\ref{fig:IC extraction}. As with Fig.~\ref{fig:CG xy chromaticity diagram}, the Planckian locus, the isotemperature lines, and the equal $\Delta$uv lines of $\pm 0.02\Delta$uv are depicted. The left side containes the saturated pixels reaching $255$, but the right side does not contain.}
    \label{fig:IC xy chromaticity diagram}
  \end{center}
\end{figure}

Applying the conversion method mentioned above, we obtained the CCT of the lightning channel.
Figs.~\ref{fig:CG xy chromaticity diagram} and \ref{fig:IC xy chromaticity diagram} denote the projected points for the CG and IC lightning channels shown in Figs.~\ref{fig:CG extraction} and \ref{fig:IC extraction}, respectively.
In Figs.~\ref{fig:CG xy chromaticity diagram} and \ref{fig:IC xy chromaticity diagram}, the Planckian locus (thick line), the isotemperature lines crossing the Planckian locus and the equal $\Delta$uv lines of $\pm 0.02\Delta$uv are also depicted.
From Figs.~\ref{fig:CG xy chromaticity diagram} and \ref{fig:IC xy chromaticity diagram}, it is found that many points distribute in the range $\pm 0.02 \Delta$uv, but there are also points distributing to the outside of the range $\pm 0.02 \Delta$uv.
Since the CCT is discussed in the range $\pm 0.02\Delta$uv, in this work we focus only on the points in the range $\pm 0.02 \Delta$uv.
The projected points for the CG lightning is mainly concentrated in the narrow CCT range, about $6000$ -- $7000$ [K], but that for the IC lightning distribute in the wide range: about $6000$ -- $50000$ [K].

\begin{figure}[hbtp]
  \begin{center}
    \includegraphics[width=60.0mm]{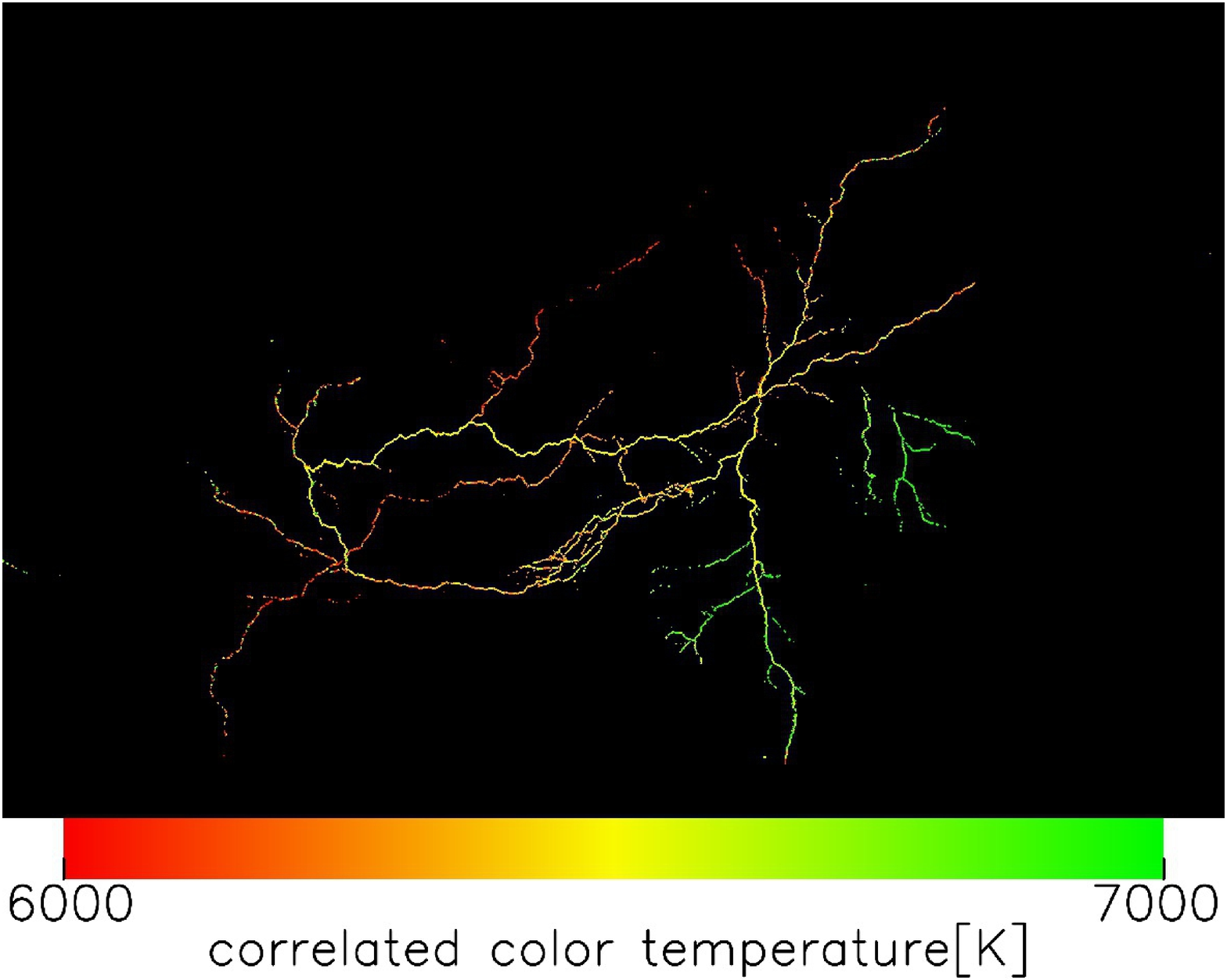}
    \includegraphics[width=60.0mm]{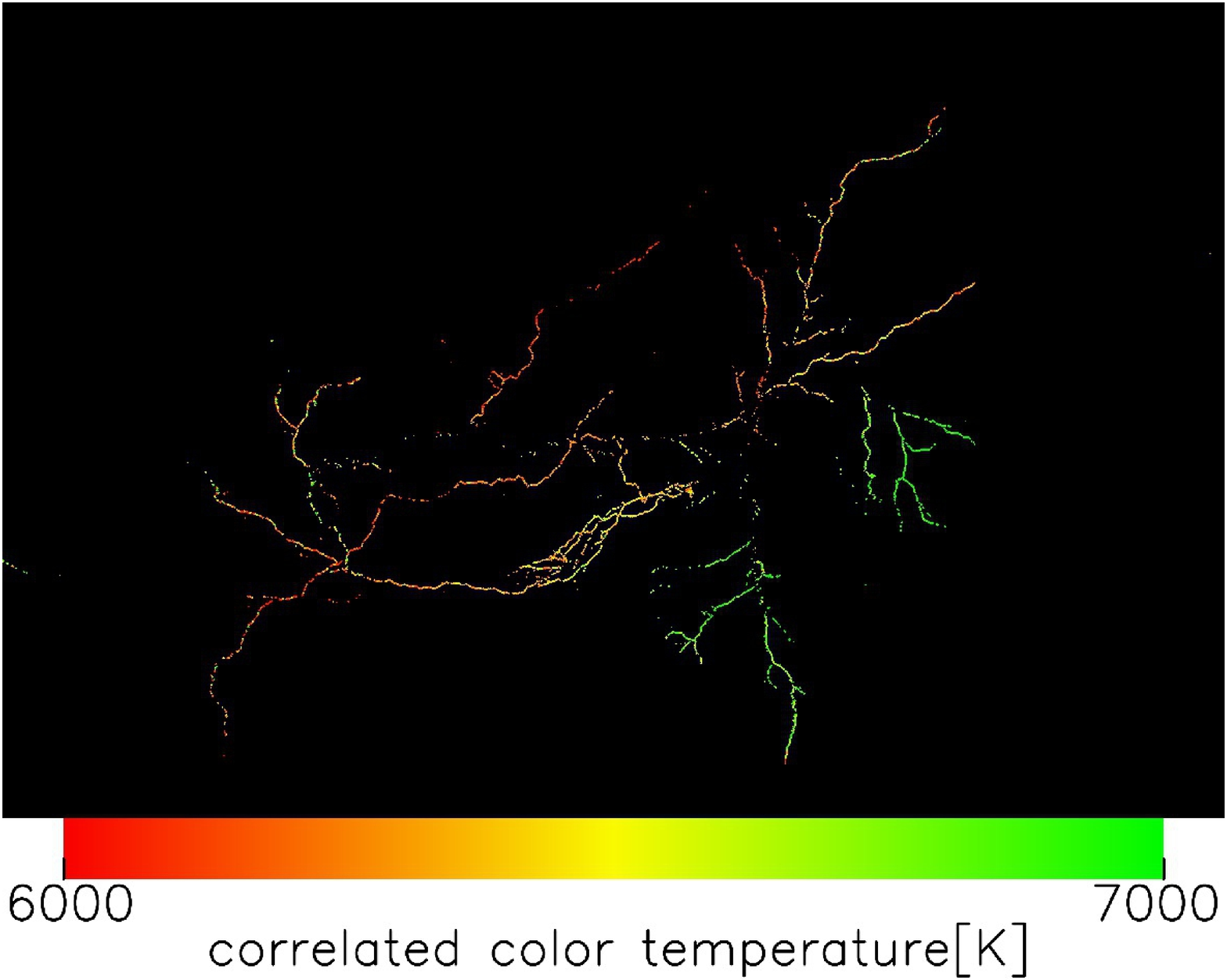}
    \caption{CG lightning images represented by the CCT. The left side containes the saturated pixels reaching $255$, but the right side does not. The lightning channel was thicken because the extracted lightning channel is very thin and confirmation of the variation of the CCT is very difficult.}
    \label{fig:CG correlated color temperature}
  \end{center}
\end{figure}

\begin{figure}[hbtp]
  \begin{center}
    \includegraphics[width=60.0mm]{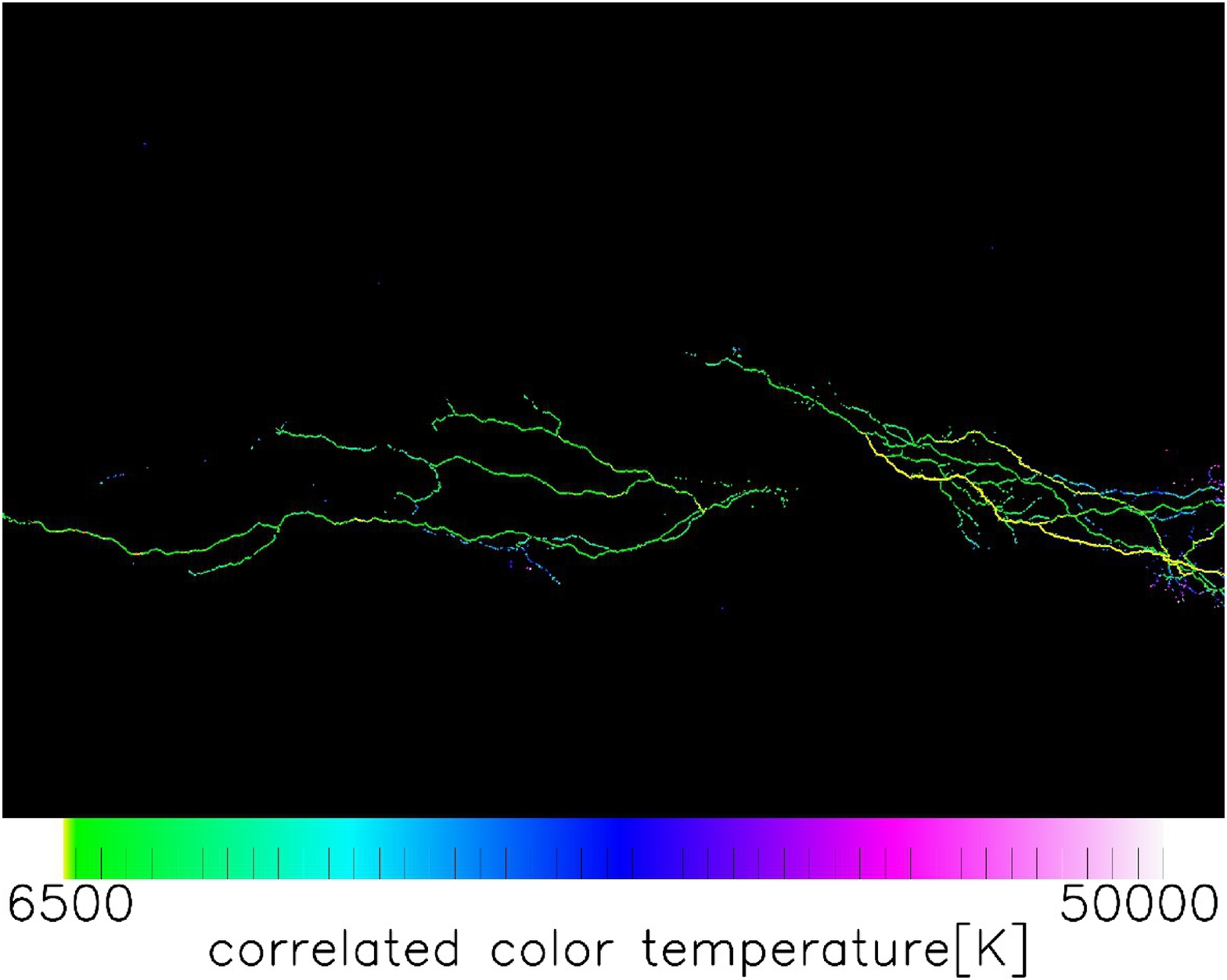}
    \includegraphics[width=60.0mm]{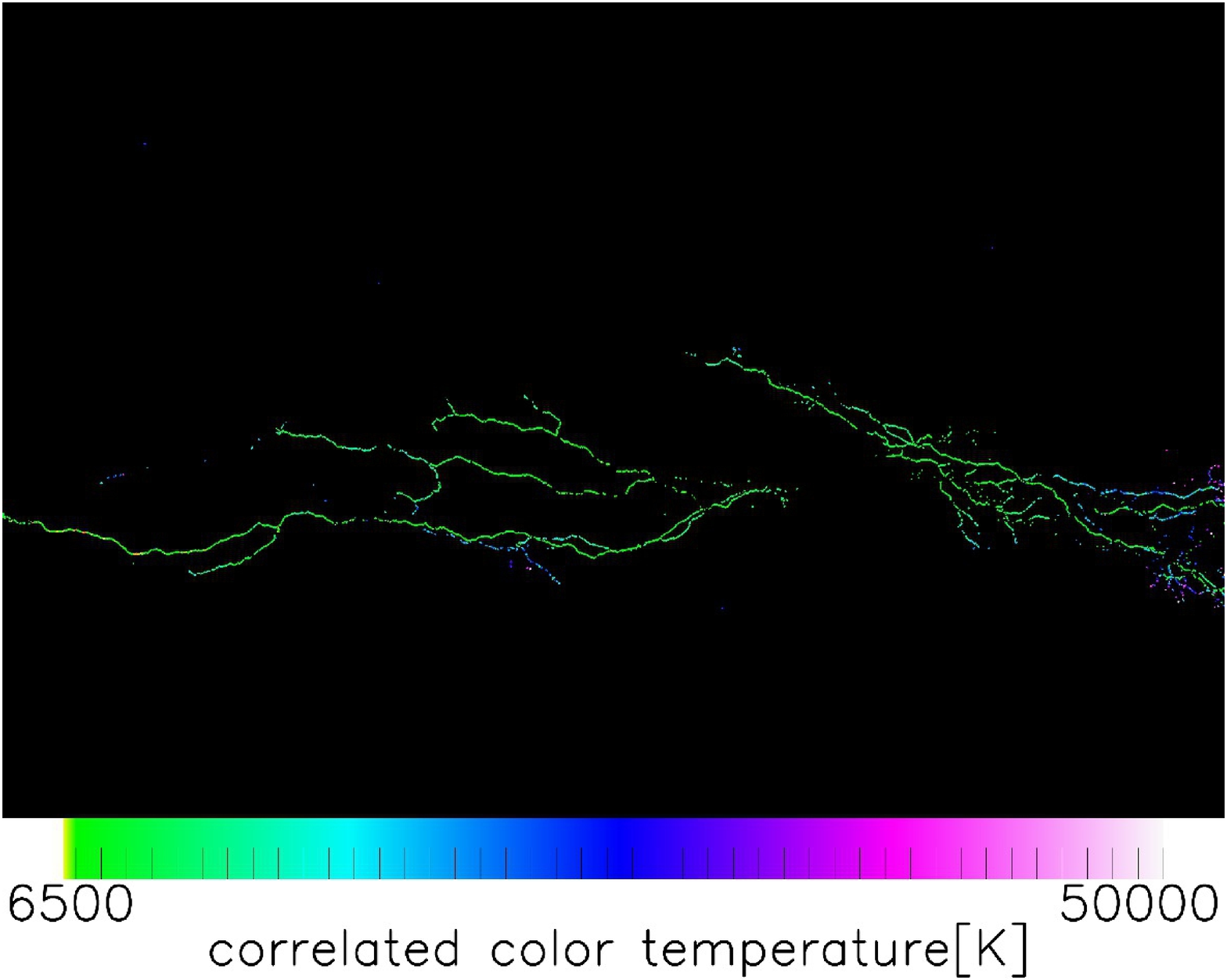}
    \caption{IC lightning images represented by the CCT. The left side contains the saturated pixels reaching $255$, but the right side does not. As with Fig.~\ref{fig:CG correlated color temperature}, the lightning channel was thicken.}
    \label{fig:IC correlated color temperature}
  \end{center}
\end{figure}

In order to look at the spatial variation of the CCT of the lightning channel, we remapped the projected points on CIE 1931 $xy$-chromaticity diagram shown in Figs.~\ref{fig:CG xy chromaticity diagram} and \ref{fig:IC xy chromaticity diagram} to the former $2$-dimensional images (see Figs.~\ref{fig:CG correlated color temperature} and \ref{fig:IC correlated color temperature}) inheriting the CCT.
The projected points within the equal $\Delta$uv lines of $\pm 0.02\Delta$uv were remapped, but the outside points of the equal $\Delta$uv lines of $\pm 0.02\Delta$uv were eliminated.
Furthermore in Figs.~\ref{fig:CG correlated color temperature} and \ref{fig:IC correlated color temperature}, to emphasize small variation of the CCT, we narrowed the CCT range to $6000$ -- $7000$ [K] for the CG flash and to $6500$ -- $50000$ [K] for the IC flash.
The color bars in Figs.~\ref{fig:CG correlated color temperature} and \ref{fig:IC correlated color temperature} indicate the CCT.
The ranges of the color bar are $6000$ -- $7000$ [K] for the CG flash and $6500$ -- $50000$ [K] for the IC flash.
Figs.~\ref{fig:CG correlated color temperature} (left) and \ref{fig:IC correlated color temperature} (left) containe the saturated pixels ($255$), but Figs.~\ref{fig:CG correlated color temperature} (right) and \ref{fig:IC correlated color temperature} (right) do not contain.
In Figs.~\ref{fig:CG correlated color temperature} and \ref{fig:IC correlated color temperature}, the lightning channel was thickened to $5$ pixels, because the extracted lightning channel is very thin, namely confirming the variation of the CCT is too difficult.
Notice that the color on the color bar is to emphasize the variation of the CCT, and does not coincide with the actual color as to the color temperature of the Planckian radiator.

%% file: sec_05.tex
\section{Discussion}
\label{sec:discussion}

\subsection{Identification of the lightning events in  the CG and IC lightning images}
\label{ssec:identification of the lightning events in the CG and IC lightning images}

\begin{table}[hbtp]
  \caption{Lightning data in 21:15 -- 22:15 JST in the investigation area in Chikusei City, Ibaraki Prefecture, Japan, on October 27th, 2008. The data was provided by Franklin Japan Corporation. The data includes ID numbers, time (JST), the location (WGS84) of lightning strike points, the polarity (+/-), the estimated peak current (kA), and CG/IC index. The latitude and longitude are given in decimal degrees. JST is expressed by $\text{JST} = \text{UTC} + 9 \ \text{hours}$.}
  \label{tbl:lightning data}
  \begin{tabular}{lllllll}
    \hline
    ID & Time (JST) & \multicolumn{2}{l}{Lightning location} & Polarity & Current & CG/IC \\
    \cline{3-4}
     & (ns)  & Latitude & Longitude & & (kA) &  \\
    \hline\hline
    1 & 21:17:6:591327450 & 36.3683$^{\circ}$N & 139.9669$^{\circ}$E & $-$ & 6 & CG \\
    2 & 21:18:27:558320409 & 36.3536$^{\circ}$N & 139.9745$^{\circ}$E & $-$ & 13 & IC \\
    3 & 21:18:27:589571036 & 36.3428$^{\circ}$N & 139.9891$^{\circ}$E & $-$ & 28 & CG \\
    4 & 21:18:27:735903133 & 36.328$^{\circ}$N & 139.9989$^{\circ}$E & $-$ & 9 & CG \\
    5 & 21:20:40:116117691 & 36.3649$^{\circ}$N & 139.9905$^{\circ}$E & $-$ & 68 & CG \\
    6 & 21:20:40:148843700 & 36.3569$^{\circ}$N & 139.9951$^{\circ}$E & $-$ & 12 & CG \\
    7 & 21:21:19:214825550 & 36.3473$^{\circ}$N & 139.978$^{\circ}$E & $+$ & 7 & CG \\
    8 & 21:22:12:826584983 & 36.3561$^{\circ}$N & 139.9834$^{\circ}$E & $-$ & 31 & CG \\
    9 & 21:22:12:849340650 & 36.3965$^{\circ}$N & 140.0426$^{\circ}$E & $-$ & 13 & CG \\
    10 & 21:23:2:212293887 & 36.334$^{\circ}$N & 140.0356$^{\circ}$E & $+$ & 14 & CG \\
    11 & 21:23:58:705277763 & 36.3595$^{\circ}$N & 140.0087$^{\circ}$E & $-$ & 11 & CG \\
    12 & 21:24:10:326889443 & 36.3608$^{\circ}$N & 139.98$^{\circ}$E & $+$ & 8 & IC \\
    13 & 21:25:18:83780450 & 36.3327$^{\circ}$N & 140.052$^{\circ}$E & $-$ & 15 & CG \\
    14 & 21:25:18:83784281 & 36.3583$^{\circ}$N & 140.0108$^{\circ}$E & $-$ & 21 & CG \\
    15 & 21:25:18:108304626 & 36.3747$^{\circ}$N & 139.9733$^{\circ}$E & $-$ & 6 & CG \\
    16 & 21:25:39:614264250 & 36.3662$^{\circ}$N & 139.9693$^{\circ}$E & $+$ & 7 & IC \\
    17 & 21:28:9:397386500 & 36.3708$^{\circ}$N & 139.9782$^{\circ}$E & $-$ & 19 & IC \\
    18 & 21:28:9:418860575 & 36.3836$^{\circ}$N & 140.077$^{\circ}$E & $-$ & 5 & CG \\
    19 & 21:29:41:912989800 & 36.3322$^{\circ}$N & 139.9863$^{\circ}$E & $+$ & 5 & IC \\
    20 & 21:32:15:530283215 & 36.316$^{\circ}$N & 140.0776$^{\circ}$E & $-$ & 35 & CG \\
    21 & 21:32:15:603968541 & 36.376$^{\circ}$N & 140.0589$^{\circ}$E & $-$ & 11 & CG \\
    22 & 21:33:12:666414500 & 36.3266$^{\circ}$N & 140.0708$^{\circ}$E & $+$ & 14 & CG \\
    23 & 21:33:12:667979150 & 36.3639$^{\circ}$N & 140.032$^{\circ}$E & $+$ & 7 & IC \\
    24 & 21:34:0:937847200 & 36.3565$^{\circ}$N & 140.0204$^{\circ}$E & $+$ & 11 & IC \\
    25 & 21:41:58:745186600 & 36.3424$^{\circ}$N & 140.0718$^{\circ}$E & $+$ & 8 & CG \\
    26 & 21:44:16:176844671 & 36.338$^{\circ}$N & 140.0767$^{\circ}$E & $-$ & 43 & CG \\
    27 & 21:44:16:187369400 & 36.3374$^{\circ}$N & 140.1224$^{\circ}$E & $-$ & 7 & CG \\
    28 & 21:46:29:646441950 & 36.3721$^{\circ}$N & 140.0536$^{\circ}$E & $+$ & 5 & IC \\
    29 & 21:46:29:847440271 & 36.4008$^{\circ}$N & 140.0779$^{\circ}$E & $+$ & 31 & CG \\
    30 & 22:3:21:489762200 & 36.3375$^{\circ}$N & 140.1234$^{\circ}$E & $-$ & 7 & IC
  \end{tabular}
\end{table}

  We have identified the strike points of the CG lightning image (Fig.~\ref{fig:lightning images} (Left)) using lightning data in Table~\ref{tbl:lightning data}.
  The data in Table~\ref{tbl:lightning data} denote the lightning events in 9:15--10:15 in the investigation area in Chikusei City, Ibaraki Prefecture, Japan, on October 27th, 2008.
The investigation area is $15$ km $\times$ $15$ km square and the center of investigation area is located at lat. $36.350173$N and long. $140.045198$E.
  The data in Table~\ref{tbl:lightning data} were commercially provided by Franklin Japan Corporation.
  The provided data were observed by the Japan Lightning Detection Network (JLDN) operated by Franklin Japan Corporation.
  The detection efficiency of the JLDN is greater than $90\%$ for CG flashes and $40\%$ for IC flashes.
  The median location accuracy is better than $0.5$ km.
  The data in Table~\ref{tbl:lightning data} include ID numbers, time (JST), the location (lat. and long.), the polarity (+/-), the estimated peak current (kA), and CG/IC index.
  The locations of lightning events are given by World Geodetic System 1984 (WGS84) and latitude and longitude are given in decimal degrees.
  The time in Table~\ref{tbl:lightning data} is based on the Japan Standard Time (JST), where using Coordinated Universal Time (UTC), JST is expressed by $\text{JST} = \text{UTC} + 9 \ \text{hours}$.

\begin{figure}
  \begin{center}
    \includegraphics[width=120.0mm]{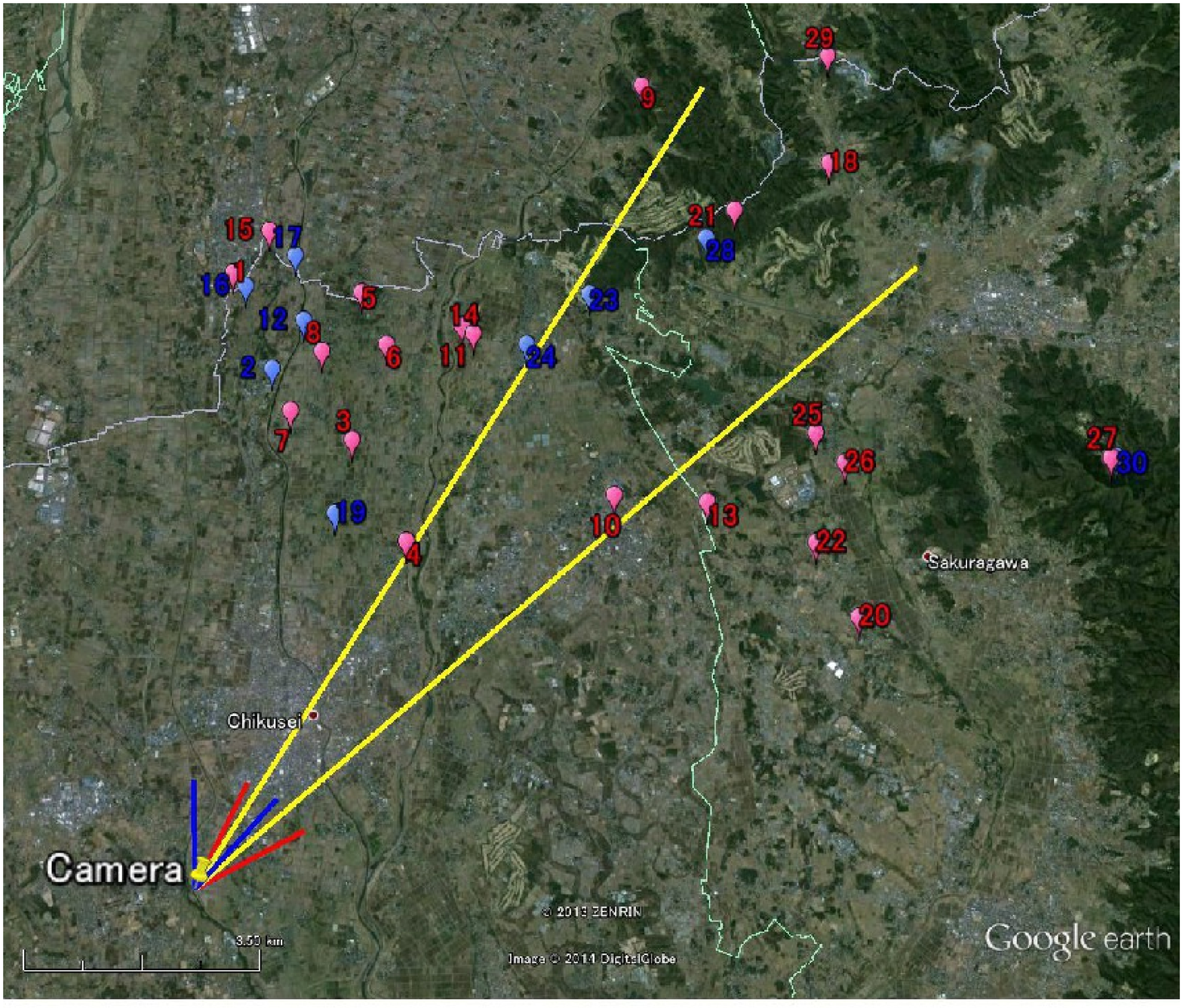}
    \caption{The locations of the lightning events in Table~\ref{tbl:lightning data}. The red balloons and the blue balloons indicate the CG and IC events, respectively. The camera location is expressed by the yellow pushpin. The red and blue lines denote the fields of view of Fig.~\ref{fig:lightning images} (Left) and (Right), respectively. The yellow lines denote the approximate direction of strike points of the CG flash. The figure was depicted using Google earth.}
    \label{fig:lightning coordinate}
  \end{center}
\end{figure}

  We have investigated the geographic locations of the lightning events in Table~\ref{tbl:lightning data}.
  Fig.~\ref{fig:lightning coordinate} shows the locations of the CG and IC lightning events in Table~\ref{tbl:lightning data} and the photographing location.
  In Fig.~\ref{fig:lightning coordinate} the CG and IC lightning events are denoted by the red and blue balloons, respectively.
  The yellow pushpin indicates the photographing location.
  Approximate directions of the two lightning strike points in Fig.~\ref{fig:lightning images} (Left) are denoted by the yellow lines.
  The red lines near the photographing location (yellow pushpin) denote the field of view of the CG flash image (Fig.~\ref{fig:lightning images} (Left)).
  The blue lines near the photographing location (yellow pushpin) are the field of view of the IC flash image (Fig.~\ref{fig:lightning images} (Right)).
  In Fig.~\ref{fig:lightning coordinate}, there are some candidates of the two strike points in the CG flash image (Fig.~\ref{fig:lightning images} (Left)) along the two yellow lines.
  Therefore it is difficult to identify the lightning strike points using only the geographical location.

\begin{figure}
  \begin{center}
    \includegraphics[width=120.0mm]{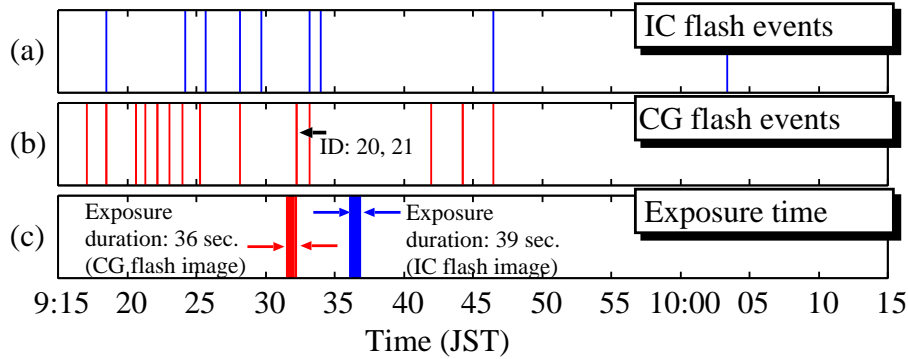}
    \caption{Schematic illustration of the time (JST) for the (a) CG and (b) IC flash events shown in Table~\ref{tbl:lightning data} and for (c) the exposure duration of Fig.~\ref{fig:lightning images}. The beginning time of the exposure is $21$:$31$:$33$ for the CG flash and is $21$:$36$:$09$ for the IC flash. The exposure duration for the CG and IC flashes are $36$ and $39$ second, respectively.}
    \label{fig:lightning events vs. time}
  \end{center}
\end{figure}

  We look at the time of the lightning events in Table~\ref{tbl:lightning data}.
  We have depicted to Fig.~\ref{fig:lightning events vs. time} the schematic illustration of the time with respect to the lightning events and the exposure duration of the CG and IC flash images (Fig.~\ref{fig:lightning images} (Left) and (Right)).
  Fig.~\ref{fig:lightning events vs. time} (a) shows the individual times of the IC flash events in Table~\ref{tbl:lightning data} and there are $9$ IC flash events in $9$:$15$ -- $10$:$15$ JST.
  Fig.~\ref{fig:lightning events vs. time} (b) shows the individual times of the CG flash events in Table~\ref{tbl:lightning data}.
  In Fig.~\ref{fig:lightning events vs. time} (b), there are $21$ lightning events in $9$:$15$ -- $10$:$15$ JST, while it seems that there are $14$ lightning events.
  Namely, seven CG flash events are overlapping because these events occurred nearly simultaneously.
  Fig.~\ref{fig:lightning events vs. time} (c) shows the exposure duration for the CG and IC flash images in Fig.~\ref{fig:lightning images} (Left) and (Right).
  The beginning time of the exposure is $21$:$31$:$33$ JST for the CG flash and is $21$:$36$:$09$ JST for the IC flash.
  The exposure duration for the CG and IC flashes are $36$ and $39$ second, respectively.
  By comparing Fig.\ref{fig:lightning events vs. time} (b) and (c), it follows that the flash events, ID: $20$ and $21$, are most appropriate to the CG flash events of Fig.~\ref{fig:lightning images} (Left) where the ID numbers are shown in Table~\ref{tbl:lightning data}.
  Thus we determined that the lightning event of the ID: $20$ is the right side of the lightning strikes in Fig.~\ref{fig:lightning images} (Left).
  Similarly the lightning event of the ID: $21$ is the left side of the lightning strikes in Fig.~\ref{fig:lightning images} (Left).

  On the other hand, we cannot identify the IC flash shown in Fig.~\ref{fig:lightning images} (Right).
  In Fig.~\ref{fig:lightning events vs. time} (a), there are no flash events near the time $21$:$36$:$09$ JST which is the beginning time of the exposure for the IC lightning events shown in Fig.~\ref{fig:lightning images} (Left).
  In the case of IC lightning, the detection efficiency of JLDN is about $40\%$.
  It is considered that the IC flash event in Fig.~\ref{fig:lightning images} (Right) could not be detected.


\subsection{Correlated color temperature and energy of the lightning channel}
\label{ssec:correlated color temperature and energy of the lightning channel}

We discuss both the CCT and related energy of the lightning channel removed the saturated pixels reaching $255$.
In general, it can be considered that the saturated part of the lightning channel have higher energy.
But we cannot know the accurate information for the saturated pixels of the lightning image.
Hence, we study the CCT and the related energy of the lightning channel removing the saturated part, namely we used Figs.~\ref{fig:CG correlated color temperature} (right) and \ref{fig:IC correlated color temperature} (right).

We think that the CCT is related to the energy of the lightning channel.
Usually, the spectra of the short wavelength increase with increasing the CCT.
Thus it can be considered that lightning channels having higher CCT have the higher energy than the other channels.
  As for the CCT of the CG lightning channel (Fig.~\ref{fig:CG correlated color temperature} (right)), we can find that the CCT around the right strike channel (ID: $20$) is relatively higher than the other channel.
  The CCT around the right strike channel (ID: $20$) distribute in about $6500$ -- $7000$ K and this CCT are displayed gradually by using the colors from yellow to green.
  The CCT around the left strike channel (ID: $21$) and other branch channels distribute in about $6000$ -- $6500$ K and are displayed gradually by using the colors from red to yellow.
  As stated above, the right strike channel (ID: $20$) have higher CCT than the others.
Thus we consider that in Fig.~\ref{fig:CG correlated color temperature} (right), the right strike channel (ID: $20$) have higher energy than the left strike channel and other channels.
Moreover, from Table~\ref{tbl:lightning data}, the currents of the left and right strick channels (ID: $21$ and $20$) are $I_{ID:21} = 11$ kA and $I_{ID:20} = 35$ kA.
Namely, $I_{\mathrm{ID:} 20} \ge I_{\mathrm{ID:} 21}$.
This means that the energy of the lightning channel having larger current is higher than that having the lower current.

  From Fig.~\ref{fig:IC correlated color temperature} (right), which was removed the saturated pixels, it is found that the CCT of the IC lightning channel mainly distribute in the range about $7000$ -- $20000$ K and is gradually displayed by the colors (green -- cyan).
  We can also find that there are the slight pixels in about $20000$ -- $50000$ K gradually displayed by the colors from cyan to blue to purple.

  From Fig.~\ref{fig:CG correlated color temperature} and \ref{fig:IC correlated color temperature}, we found that the CCT of the lightning channel changes spatially.
  This also means that the energy of the lightning channel changes spatially.
Totally, the CCT on the IC lightning channel is higher than the CG lightning channels.
This indicate that the energy of the IC lightning channel shown in Fig.\ref{fig:lightning images} (Right) was higher than the CG lightning channel shown in Fig.\ref{fig:lightning images} (Left).

  We add the discussion for the saturated pixels on the lightning channel.
  In Fig.~\ref{fig:CG correlated color temperature} (left) and \ref{fig:IC correlated color temperature} (left) the yellow channels having 6500 K (disappear in Fig.~\ref{fig:CG correlated color temperature} (right) and \ref{fig:IC correlated color temperature} (right), respectively) are the saturated channels.
  The corresponding channels in the original images shown in Fig.~\ref{fig:lightning images} (Left) and (Right) are seen more brightly.
  It is reported by Idone and Orville~\cite{Idone_Orville}, Gomes and Cooray~\cite{Gomes_Cooray}, Wang et al.~\cite{Wang}, and Zhou et al.~\cite{Zhou_et_al} that there exists a strong positive correlation between the channel current and light intensity of lightning channel.
  For the above reason, it can be considered that the current on the saturated part is high, although the details on the saturated part cannot be known.
  It can be summarized that the bright channel on the saturated part have high current and high energy.
  Furthermore, Rubenstei et al.\cite{Rubenstein_et_al} and Rakov et al.\cite{Rakov_et_al} reported that there is a positive linear correlation between the peak current and the leader electric field change.
  It is considered that the electric field changes around the bright channel having the saturated pixels is large.

%% file: sec_06.tex
\section{Conclusions}
\label{sec:conclusiosn}

  From Fig.\ref{fig:CG correlated color temperature} and \ref{fig:IC correlated color temperature}, we have found that the CCT of the lightning channel changes spatially.
  This also means that the energy of the lightning channel changes spatially.
  Since the lightning images (Fig.\ref{fig:lightning images} (Left) and (Right)) were captured by exposing the image sensor, the amount of light captured by the image sensors in the still camera was integrated in the exposure time $36$ second for the CG flash and $39$ second for the IC flash.
  For this reason, we did not analyze the temporal variation.
  If high-speed video camera is used, the more details of the spatio-temporal variation of both the CCT and energy of the lightning channel will be obtained.

%% file: sec_07.tex
\section{Acknowledgements}
\label{lbl:acknowledgements}

The authors would like to gratefully acknowledge storm chaser Yutaka Aoki for providing the digital still images of the CG and IC flash and would also need to acknowledge Franklin Japan Corporation that commercially provided JLDN data.
The authors would like to thank Yu Iida and Singo Sakihama for valuable discussion.